\documentclass[acmsmall]{acmart}
\AtBeginDocument{%
  }

\setcopyright{acmlicensed}
\acmJournal{PACMHCI}
\acmYear{2025} \acmVolume{9} \acmNumber{2}
\acmArticle{CSCW040} \acmMonth{4}
\acmDOI{10.1145/3710938}

\usepackage{multirow}

\usepackage{csquotes}

\usepackage{graphicx}
\usepackage{subcaption}
\graphicspath{{images/}}

\begin{document}

\title[BoundarEase]{BoundarEase: Fostering Constructive Community Engagement to Inform More Equitable Student Assignment Policies}

\author{Cassandra Overney}
\affiliation{%
  \institution{Massachusetts Institute of Technology}
  \city{Cambridge}
  \state{Massachusetts}
  \country{United States}}
\email{coverney@mit.edu}

\author{Cassandra Moe}
\affiliation{%
  \institution{Northeastern University}
  \city{Boston}
  \state{Massachusetts}
  \country{United States}}
\email{moe.c@northeastern.edu}

\author{Alvin Chang}
\affiliation{%
  \institution{The New School}
  \city{New York}
  \state{New York}
  \country{United States}}
\email{alvinchang@newschool.edu}

\author{Nabeel Gillani}
\affiliation{%
  \institution{Northeastern University}
  \city{Boston}
  \state{Massachusetts}
  \country{United States}}
\email{n.gillani@northeastern.edu}

\renewcommand{\shortauthors}{Cassandra Overney, Cassandra Moe, Alvin Chang, and Nabeel Gillani}

\begin{abstract}
Public school districts across the United States (US) play a pivotal role in shaping access to quality education through their student assignment policies---most prominently, school attendance boundaries.  Community engagement processes for changing such policies, however, are often opaque, cumbersome, and highly polarizing---hampering equitable access to quality schools in ways that can perpetuate disparities in achievement and future life outcomes.  In this paper, we describe a collaboration with a large US public school district serving nearly 150,000 students to design and evaluate a new sociotechnical system, ``BoundarEase'', for fostering more constructive community engagement around changing school attendance boundaries.  Through a formative study with 16 community members, we first identify several frictions in existing community engagement processes during boundary planning, like individualistic over collective thinking; a failure to understand and empathize with different community members when considering policy impacts; and challenges in accessing and understanding the impacts of boundary changes.  We then use these frictions to inspire the design and development of BoundarEase, a web platform that allows community members to explore and offer feedback on potential boundaries based on their preferences.  A user study with 12 community members reveals that BoundarEase prompts reflection among community members on how policies might impact families beyond their own, and increases transparency around the details of policy proposals.  Our paper offers education researchers insights into the challenges and opportunities involved in community engagement for designing student assignment policies; human-computer interaction researchers a case study of how new sociotechnical systems might help mitigate polarization in local policymaking; and school districts a practical tool they might use to facilitate community engagement to foster more equitable student assignment policies.
\end{abstract}



\received{January 2024}
\received[revised]{July 2024}
\received[accepted]{October 2024}

\maketitle

\section{Introduction}
Across the US, student assignment policies play a foundational role in enabling, or impeding, access to quality public K-12 education. The policies school districts adopt can impact the opportunities and resources students have access to at their schools~\cite{monarrez2021urban}; levels of racial/ethnic and socioeconomic integration~\cite{monarrez2023boundaries}---which in turn can affect students' and parents' access to social capital and other valuable network connections~\cite{chetty2022socialcapitalII,smallUnanticipatedGains}; levels of school utilization~\cite{hawkins2023utilization}; and several other factors.

One policy regime, in particular, continues to influence how the vast majority of students across the US are assigned to schools: school ``attendance boundaries''.  Attendance boundaries are catchment areas that districts draw to determine which neighborhoods should feed to which schools.  Typically, students are assigned to the schools closest to their homes~\cite{monarrez2023boundaries}. This often means school assignments reflect the underlying geographic inequities in American communities~\cite{monarrez2023boundaries}.  Attendance boundaries are powerful default settings: as of 2016, 71\% of students attended their (presumably neighborhood-assigned) schools~\cite{nces2021choicefacts}.  This shows that, even in an era of growing school choice~\cite{houlgrave2021choice}, attendance boundaries continue to play a critical role in shaping the quality of education that students are able to access.  

Thus, changing attendance boundaries (``redistricting'') may serve as a powerful lever for advancing more equitable access to quality education---especially for students of color, students living in poverty, 
and other historically disadvantaged groups.  Indeed, a recent study showed that changes to elementary school boundaries across a set of large US districts could induce moderate reductions in racial/ethnic segregation while also slightly reducing travel times~\cite{gillani2023redrawing}.  Furthermore, some estimates suggest that attendance boundary changes are relatively common across the US, with approximately 15\% of boundaries, annually, experiencing some type of change\footnote{\url{https://www.attomdata.com/data/boundaries-data/school-attendance-zone-boundaries/}.}.  Still, changing attendance boundaries is highly contentious, particularly when doing so involves efforts to promote more equitable student assignments: such efforts can result in superintendent resignations~\cite{bridges2016eden}, family flight to other school options~\cite{mervosh2021minneapolis}, and intense community strife~\cite{genevieve2017rezoning,castro2022richmond,bridges2016eden}.  Furthermore, such contention---particularly in the present day---happens against a backdrop of rising political polarization, with communities split on which books their children should be reading in school~\cite{smith2023books}, how reading should be taught~\cite{wilkins2023reading}, and a host of other factors.

Given these challenges, how might we practically alter attendance boundaries to sustainably achieve more equitable student assignment policies?  Exploring this question first requires defining what we mean by ``equity''.  The notion of equity, of course, has many definitions.  A focus on increasing access to quality ``goods'' (in this case, schools), particularly among populations who lack such access, reflects elements of Rawlsian equity/justice~\cite{sen1979equality}.  This is the notion we seek to advance throughout this study: student assignment policies have long relegated historically disadvantaged groups into segregated learning environments, which often lack the privileges and amenities of schools in more affluent areas.  Parent power and preferences tend to reinforce these patterns and further reduce access to quality education among other groups.  Addressing these inequities sustainably, however, also requires addressing inequities in access to information: families may have very different understandings of how student assignment policies work, and what their options under such policies might be~\cite{robertson2021vsd}.  We view addressing these informational inequities as an important part of fostering more equitable access to quality schools. 

Importantly, Rawlsian equity suffers from many limitations---including the fact that it does not account for the unique supports that different sub-populations may require to take full advantage of goods once they are able to access them~\cite{sen1979equality}.  While it is incomplete, Rawlsian equity offers a useful lens for framing what a high-level objective of student assignment policy-making might be.  However, if we wish to advance more equitable policy-making, it is worth first understanding some of the forces that currently impede it in practice.  As described in~\cite{gillani2023air}, one impediment is that the loudest voices during a redistricting process are often the ones that get the most attention~\cite{genevieve2017rezoning,bridges2022comment}, drowning out others that may help moderate an otherwise polarized process.  Another is transparency: families can thwart redistricting efforts by challenging their legitimacy, often citing a lack of transparency in the district's policymaking and community engagement processes.  A more deeply entrenched impediment is prioritizing the individual over the collective---a tendency that is often driven by fear.  Families may fear a drop in home values if their zoned schools change~\cite{bridges2016eden,kane2005housing}; fear that their children won't receive a quality education at their new school~\cite{zhang2008flight}; etc.  These fears, while sometimes legitimate, can block progress and lead families to exclusively support policies that benefit themselves, even if small amounts of compromise might produce vastly improved outcomes for others---creating a ``tragedy of the commons'' that can be difficult to escape~\cite{cornwall2017tragedy}.  Finally, families may exhibit racialized preferences for schools~\cite{billingham2016parents,hailey2021parents}, favoring boundaries that put their children in schools with certain peer groups over others.

These are complex sociopolitical forces---and some, particularly the last one (racialized preferences)---are likely to require a longer time horizon for change.  They represent human challenges that may seem too formidable for data and technology-enabled tools to help address.  Yet as~\cite{gillani2023air} suggests, thoughtfully designed sociotechnical systems may have a role to play in mitigating them.  For example, such systems may help broaden the set of voices that participate in community engagement processes for redistricting efforts, diluting the loud voices that often dominate policymaking discussions.  They may lucidly illustrate the district's policymaking process and the impacts proposed boundaries might have on individual families, helping to combat challenges to transparency.  Sociotechnical systems might also help families more clearly see how a rezoning can impact \textit{other families} in the district beyond their own---perhaps prompting people to consider how their preferences might affect others in both positive and harmful ways.

In this paper, we ask: how might we design sociotechnical systems to help achieve some of the above possibilities to foster support for attendance boundary changes that might expand equitable access to quality education?  To explore this, we collaborate with a large US public school district serving nearly 150,000 students. The district has previously faced challenges in engaging diverse populations around consequential policy decisions, typically using lengthy PowerPoint presentations to update families on boundary proposals, which has made it difficult for those with little time or background context to meaningfully shape the policymaking process.

In collaboration with the district, we conduct 16 formative interviews with parents involved in a recent, large high school redistricting process affecting over 10,000 families to better understand their perceptions of, and experiences with, the rezoning process.  From the formative interviews, we distill a number of themes that inform design questions for a new sociotechnical system---``BoundarEase''---to help districts more effectively facilitate community engagement in pursuit of equitable attendance boundary design.  These questions concern fostering both individual and collective thinking, exposing community members to the perspectives of those different from them (``perspective-getting''~\cite{kalla2021perspective}), and increasing understanding of the expected impacts of a rezoning by using a shared language and standard framework through which to present them.

We evaluate BoundarEase's potential to help districts address these questions by conducting a user study with 12 community members.  The study shows that BoundarEase appears to prompt reflections among participants on how boundary changes might impact families beyond their own---a necessary, though not sufficient, first step towards fostering more collective thinking.  BoundarEase makes less progress on a feature designed to prompt perspective-getting, with mixed responses from community members.  Participants indicate the most progress on BoundarEase's presentation of proposed boundaries, saying that it brings clarity and transparency to an otherwise convoluted process of trying to understand what personal and collective impacts boundary changes might have.  We release all of the software, algorithms, and data developed for this study in the supplementary materials to support both researchers and practitioners in building on this work\footnote{\url{https://github.com/Plural-Connections/BoundarEase-platform}.}.  

Our study offers education researchers a more nuanced understanding of the challenges and opportunities surrounding redistricting-related community engagement; human-computer interaction researchers a case study of how sociotechnical systems might be designed to help mitigate polarization in local policymaking; and school districts a practical tool they might use to facilitate community engagement to foster more equitable student assignment policies.

\subsection{Positionality Statement}
Our backgrounds span a number of disciplines, including human-computer interaction, data science, journalism, and design.  We have shared interests and experiences in designing new tools and systems to help mediate contentious conversations---particularly ones that tend to occur in person.  Some of the authors have previously collaborated with the school district on another project, and through that effort, had the opportunity to build a strong working relationship with our district sponsor prior to the start of this study.  Some of the authors have also previously explored methods for fostering more diverse and integrated schools.  While school diversity is certainly a relevant topic here, we emphasize a broader aim of this study: to build bridges between community members who may have very different priorities and visions of what a desirable student assignment policy entails (including community members who may not value diversity at all).  This aim has caused us to ensure that our own individual policy desiderata do not unduly influence our research or design process, and that our approach ultimately advances the broader goal of fostering more equitable student assignment policies instead of any single policy objective.

\section{Related Work}

\subsection{Student Assignment Policymaking and Community Engagement}
School districts typically devise and deploy a myriad of policies to collectively shape students' assignments to schools.  Within districts, families may opt for ``choice'' programs that enable their children to attend magnet or otherwise specialized offerings.  Families may also opt out of assigned public schools entirely, choosing public charter schools or private/parochial options---and, indeed, sometimes opt out \textit{because} of districts' student assignment policies~\cite{mervosh2021minneapolis}.  This makes student assignment policymaking a critical function of local education officials: one that can dramatically alter intergenerational life outcomes for students and families.

Still, student assignment policymaking---and attendance boundary changes in particular---continues to be a highly contentious topic.  Most of the literature on the topic spans both academic and non-academic sources---and often highlights the impediments communities and districts face in passing changes that promote equity.  For example, a case study by the Century Foundation of Eden Prairie, Minnesota, described the intense backlash district officials faced when proposing boundary changes that would help increase access to better-resourced learning environments for lower-income populations of color.  While the boundaries passed, most of the school board, including the Superintendent, resigned in response to public pressure~\cite{bridges2016eden}.  This backlash is captured through racially-coded community feedback shared in other places~\cite{castro2022richmond, castro2022rezoning, mendez2022williamsburg}.  The opposition is often magnified when a subset of families---often those with the confidence to speak publicly, and with both the time and resources to mobilize like-minded parents---speak out against policies that they believe may harm their children relative to others~\cite{genevieve2017rezoning}.  This imbalance in participation can drown out more moderate and minority voices, yielding a form of ``public comment inequity''~\cite{bridges2022comment} and hampering the creation of equitable student assignment policies that meet families where they are~\cite{robertson2021vsd,robertson2022school}.

Public input processes that favor the loud and well-resourced often result in policies and tools that fail to meet the needs of historically marginalized community members who, in turn, may opt out of policy feedback channels. These failure modes prevent public educational agencies from fulfilling their promise to deliver quality education to all.   It is unclear, however, to what extent districts mitigate these failure modes, as designing effective and inclusive community engagement paradigms continue to pose a challenge across educational policymaking settings~\cite{makori2021districts}.  
These shortcomings underscore the need for new explorations into how districts can design and execute community engagement initiatives that advance more equitable student assignment policies.  Some researchers have proposed such explorations, including models for how computers might support humans in ``wicked'', socio-politically complex activities like boundary planning~\cite{sistrunkCSCWAmpRedrawing2023}.  We aim to build on these efforts in our study with the design and preliminary evaluation of a specific system for community-based attendance boundary planning.

\subsection{Collective Versus Individual Thinking}
Given how contentious student assignment policymaking can be, a natural question is: why?  The paper's introduction offers some possible causes---one of which involves a tendency for many, particularly in Western countries (like the US), to prioritize individual freedom and gain over collective well-being~\cite{buss2000evolution}.  Indeed, this played out during the COVID-19 pandemic, with research showing that those living in regions that exhibited a greater sense of collectivism were also more likely to wear masks~\cite{lu2021masks}.  

Individualistic thinking is not confined to a specific political leaning.  For example, adopting a ``not-in-my-backyard'' mindset~\cite{hankinson2018renters}, families in progressive communities have blocked policies that might advance collective access to scarce resources like affordable housing, particularly when such policies might translate into an individually sub-optimal result---like the fear of single-family home property values dropping if neighborhoods build more affordable multi-family units~\cite{kahlenberg2023excluded}.  This individual optimization also plays out in public education: through boundaries, families essentially obtain a public good (access to public schools) through a private channel (purchasing homes in desirable school catchment areas)---sparking strong opposition to changes that could affect their children's access to schools, even if such changes increase access to quality options for others.  A rigid prioritization of individual over collective well-being can yield ``tragedies of the commons,'' where finite, shared resources are depleted due to repeated individually focused optimizations.  We see such tragedies in the fight against climate change~\cite{ostrom1990governing} and in education~\cite{cornwall2017tragedy}.  

Fostering more consideration of the collective to drive more socially conscious policymaking is, of course, challenging.  
At its core, collective thinking requires conceptualizing oneself as a single node in a broader network of people---all of which are worthy and deserving of a quality life.  
Some sociologists have highlighted how conferring status and other social rewards upon individuals who take action on behalf of the collective may encourage more collective optimization~\cite{willer2009groups}.  
Yet status promotion as a reward for collective action may fail to advance the broader collective's well-being in highly polarized contexts: status may accrue precisely because people are promoting collective action \textit{in their own group}, and often, against others. 
How might we promote consideration, and activities that further the well-being, of the broader collective---which may include groups of individuals with opposing worldviews, political perspectives, and life experiences?  
Here, work by social psychologists on perspective-taking and perspective-getting may help.  Both perspective-taking and perspective-getting broadly refer to the practice of ``stepping into someone else's shoes'' to better understand their life experiences and worldviews~\cite{kalla2021perspective}. 
Perspective-taking involves \textit{imagining} an experience from another person's perspective~\cite{galinsky2000perspective}, while perspective-getting involves \textit{hearing} another person's point of view through thoughts, feelings, and experiences~\cite{eyal2018perspective}.
Perspective-taking is self-generated by an individual, while perspective-getting is received from another.  
Perspective-getting has been shown to advance connection and inclusion across disparate groups---for example, promoting more inclusionary behaviors towards refugees~\cite{adida2018perspective}, among others.  
Recent studies like~\cite{saveski2022perspective} have also highlighted how digital interfaces might scaffold perspective-taking and perspective-getting in contentious and polarized settings, in part by helping people more easily explore and reason about why others might hold viewpoints that differ from their own (i.e., promote ``cognitive empathy''~\cite{selman1980growth}).  

A few human computer interaction (HCI) systems for civic decision-making take inspiration from perspective-taking.
Bowen et al. created Your Journeys, a website that prompted users to consider how changes to public transport would fit the needs of other people.
Users followed six personas and then rated various design features from a persona's point of view~\cite{bowenMetroFutures20202023}.
In PolicyScape, users were asked to think about a policy from the perspectives of other stakeholders and to compare their guesses with actual responses~\cite{kimCrowdsourcingPerspectivesPublic2019}.
In terms of sharing diverse perspectives through perspective-getting, Faridani et al. designed interactive cluster visualizations to enable users to explore a diversity of comments, which led to higher levels of agreement and respect compared to listing comments in chronological order~\cite{faridani2010opinion}.
Other systems have implemented algorithms to expose users to different points of view.
The California Report Card platform uses an uncertainty-minimizing sampling algorithm to select suggestions from individuals who hold similar and dissimilar positions~\cite{nelimarkkaComparingThreeOnline2014}.
The Living Voters Guide encourages people to explore and clarify pro and con arguments from people with different stances on an issue or from varying segments of a population~\cite{kripleanSupportingReflectivePublic2012, kripleanThisWhatYou2012}. 
Of course, perspective-taking and perspective-getting are not the only approaches to fostering more collective thinking and consideration, but they offer a promising channel that may have applications in student assignment policymaking.  
More broadly, we believe the design and development of human-computer interfaces for fostering greater collective thinking represents an under-explored, yet rich area of future research for the computer-supported cooperative work (CSCW) community.

\subsection{HCI Systems for Civic Decision-Making}

Numerous HCI systems have been designed to support civic decision-making across diverse application areas, including urban planning and policymaking.   
These systems play a pivotal role in fostering community engagement and shaping the dialogue between community members and decision-makers.
However, there is a lack of HCI systems that support policymaking efforts in school districts, specifically for contentious student assignment changes.
BoundarEase is a novel platform designed to address this gap. 
In this section, we draw insights from prior work in the HCI and civic decision-making space, which we build upon through BoundarEase. 


To reach higher levels of citizen participation, civic platforms must establish a two-way information flow between decision-makers and community members.
Various platforms have improved the dissemination of information from decision-makers to the public, aiding residents in comprehending the context of complex decisions in areas such as voting patterns~\cite{kinnaird2010connect, alvarez2014voting}, budgeting~\cite{kimFactfulEngagingTaxpayers2015}, public transport consultations~\cite{bowenMetroFutures20202023}, policymaking~\cite{bojovicOnlineParticipationClimate2015} and urban planning~\cite{tianParticipatoryEplanningModel2023, haaslyonsExploringRegionalFutures2014}. 
Notable features include providing simulation capabilities~\cite{haaslyonsExploringRegionalFutures2014, tianParticipatoryEplanningModel2023}, supporting multi-criteria analysis~\cite{bojovicOnlineParticipationClimate2015}, creating virtual environments with game elements~\cite{gordon2011playing}, and graphically depicting the impacts of policy choices~\cite{haaslyonsExploringRegionalFutures2014}.
BoundarEase builds upon this work by providing a comprehensive explanation of the potential impacts of boundary changes, allowing community members to simulate different student assignment scenarios produced through an optimization algorithm that builds on~\cite{gillani2023redrawing}.

Regarding the information flow from the community back to decision-makers, many platforms assist in collecting feedback synchronously or asynchronously.
Prior work has created systems that improve the collection of feedback during in-person or virtual town halls~\cite{jasimCommunityClickCapturingReporting2021, jasimCommunityClickVirtualMultiModalInteractions2023}.
In-person deliberation tools combine online and physical affordances including hybrid tabletops~\cite{johnsonCommunityConversationalSupporting2017} and multiple displays~\cite{mahyarUDCoSpacesTableCentred2016}. 
Other efforts have explored ways to create situated digital technology where people could give feedback at designated locations within their community and view visualizations to see how their responses compare with their peers'~\cite{koemanEveryoneTalkingIt2015, fredericksDigitalPopUpInvestigating2015, schiavoAgora2EnhancingCivic2013, golsteijnSensUsDesigningInnovative2016, valkanovaMyPositionSparkingCivic2014, schroeterEngagingNewDigital2012}.

Not everyone can or feels comfortable with sharing their feedback in a public location, so it is essential to establish channels for individuals to provide feedback asynchronously from any location, a capability supported by online platforms. 
Gün et al. examined 25 information and communications technology-based participation platforms for urban planning and observed several common functions including adding placemarks on maps, tagging content, and voting and ranking options~\cite{gunUrbanDesignEmpowerment2020}.
Map features, particularly in the context of spatial planning~\cite{nuojuaWebMapMediaMapbasedWeb2010, yuFacilitatingParticipatoryDecisionmaking2009}—including school attendance boundary changes—are invaluable. 
Though interactive maps are powerful, they can be difficult to use, particularly for individuals lacking a technical background~\cite{rzeszewski2019usability, unrau2019usability}. 
In our formative study, we identified several challenges associated with using maps to comprehend boundary changes.
These challenges are addressed in BoundarEase through the creation of a simple and mobile-friendly map interface.


Asynchronous online platforms empower the public to gain insights into diverse perspectives and engage in deliberation.
Existing civic technologies have explored a variety of methods to transparently share community feedback.
Some platforms are fully transparent with community feedback, displaying everyone's contributions~\cite{mayDesignCivicTechnology2018, maskellSpokespeopleExploringRoutes2018}.
Regan et al. implemented familiar visualization methods, such as pie charts and bar graphs, to visualize community sensing and voting data~\cite{regan2015designing}.  
Other platforms draw on principles from crowdsourcing to encourage community collaboration for policymaking~\cite{aitamurtoFiveDesignPrinciples2015, kimCrowdsourcingPerspectivesPublic2019, kleinHowHarvestCollective2007} and urban planning~\cite{mahyarCommunityCritInvitingPublic2018, ledantec2015planning}.
BoundarEase took inspiration from several features within the Living Voters Guide, including framing interactions around pro and con considerations, representing stance through a spectrum instead of a binary yes or no answer, and prompting users to reflect on their stance while using the platform~\cite{kripleanSupportingReflectivePublic2012}.
In the realm of HCI systems for civic decision-making, BoundarEase is novel because it focuses on policymaking within public school districts—a domain that has been underexplored in prior research. 
Nelimarkka reviewed systems that support democratic decision-making and identified a need for systems that support decision-making in representative-type political contexts~\cite{nelimarkka2019review}. 
BoundarEase addresses this gap by facilitating community engagement through the design of human-data interactions in a large US public school district, where elected school board members shape student assignment policies, often after seeking feedback from community members.


\section{Formative Study}

To understand the challenges around participating in community engagement initiatives for school attendance boundary planning, we conducted a formative interview study with parents who were actively going through a boundary change.

\subsection{Research Setting}

Building upon prior in-the-wild HCI research~\cite{rogers2017research}, we developed a platform intimately tied to a real-world school boundary change process.  
Consequently, we designed and evaluated BoundarEase within a prominent school district: one of the top 20 largest in the United States—serving approximately 150,000 students across more than 180 schools, where approximately 96\% of schools rely on attendance boundaries as a pivotal element of their student assignment policy.

Our study focused on a specific region of the school district that was undergoing significant changes in school attendance boundaries.
Nestled in the southern part of the district, this region 
serves around 10,000 families. 
The area originally had four high schools.
Due to rapid population growth, the high schools were starting to overcrowd, so the School Board decided to create a new relief high school.
As a result, the school district embarked on a comprehensive redrawing of attendance boundaries for all high schools in the affected area. According to board policy, the new plan was required to strike a balance across four policy pillars: socioeconomic (SES) diversity (having the distribution of low, medium, and high-SES students at each school be as close to the district-level proportions of 1/3, 1/3, 1/3 as possible---a measure of evenness that is related to the dissimilarity index of segregation~\cite{massey1988dimensions}); home-to-school distance (ensuring students are not assigned to schools that are very far away); school utilization (to ensure schools are not too over- or under-enrolled); and the stability of feeder patterns (i.e., reducing cases where students attending the same elementary school are then split between different middle schools, and similarly for middle-to-high).
This challenging high school redistricting task occurred simultaneously with middle and elementary school redistricting, all under very tight timelines and with limited human resources in the district's Student Assignment Team.

Before finalizing the attendance boundaries, the school district was required to obtain feedback from the community.
The Student Assignment Team implemented a 15-month-long community engagement process, encompassing 18 boundary proposals that were shared with the public for feedback. 
This included surveys, community meetings, and a parent working group.
In total, they received around 10,000 feedback entries with at least 1,000 participants attending in-person and virtual engagements.  Yet engagement did not come without challenges, particularly in how families accessed and made sense of information.  For example, the district typically updated families on the latest boundary proposals using lengthy PowerPoint presentations, making it difficult for those with little time or background context to stay afloat on updates and meaningfully shape successive policy waves.  
%
Following the community engagement phase, the Student Assignment Team developed a final recommendation, which was presented to the Superintendent. The Superintendent shared the proposal with the School Board for a vote, ultimately securing approval.
\subsection{Formative Study Method}


We conducted semi-structured interviews with 16 parents who 
engaged in the recent boundary change described above.
To recruit participants, the Student Assignment Team reached out to members of the parent working group, which had at least one parent representative from each school in the impacted area. Members of the parent working group shared the interview opportunity with parents in their schools.
At the time of the formative study, the boundary change was in its final stages and a few months away from School Board approval. 
We refer to the participants from F1 to F16.
Five participants were part of the parent working group, and four participants said they were less active in the community engagement process.  
All except one participant is female with eleven living in high socioeconomic areas and five living in medium socioeconomic areas. 
Participants had children who attended grades ranging from preschool to 12th grade.
Thirteen participants had children attending multiple schools.  
In total, we talked with people who represented 14 out of 40 schools in the impacted area.
More participant details are in Table~\ref{tab:formativeparticipants} in the Appendix.
We provided \$30 Amazon gift cards to participants for an hour-long conversation.
The interview protocol received Institutional Review Board approval.

We began each interview by asking several background questions about participant involvement in the ongoing boundary change.
Then, we gave participants 20 minutes to interact with a map interface that the school district previously created.
The district's map interface was built using geographic information system (ArcGIS) software and consisted of an interactive map with several layers that users could toggle on and off through a sidebar.  
The layers included possible high school boundary proposals; existing high school, middle school, and elementary school boundaries; and SES data for each census block in the area.
We asked participants to give feedback on a boundary proposal while exploring the map. 
Afterward, participants were invited to reflect on the ongoing community engagement process. 
See supplementary materials for the questions we asked.
%
Interviews were conducted remotely over video calls, which were recorded and transcribed using an auto-transcription service.
Then, the first two authors went through the transcripts and coded them for themes using affinity diagramming, a popular analysis method in user interface design~\cite{lucero2015using} where text data is segmented into notes and then clustered into groups. 
Through multiple iterations, along with periodic discussions with the rest of the research team and the Director of the Student Assignment Team, we identified several user challenges and subsequently the design questions for our platform, BoundarEase.

\subsection{Findings}

We identified several themes from the formative study, which we highlight in the following sections.

\subsubsection{Usability Challenges in District's Current Process}\label{finding_usability}

All participants emphasized how cumbersome the existing community engagement process is.
It takes a significant amount of time to find information about the redistricting.
As F6 stated, \emph{``with the current website, information is usually outdated and hard to find.''}
In addition, information is not centralized in one location, and the \emph{``website is hard to navigate''} (F8).
F14 said that engagement opportunities \emph{``need to be better publicized. No one is fishing around on the website for updates.''}
When updates are shared, parents don't receive enough time to give feedback: \emph{``[The school district] only left a week for parents to get information and vote, so the only way parents can make change is to be really loud. I want more time to be involved''} (F1).
Besides the overall engagement process, eight participants expressed difficulties in using the map interface to understand boundary scenarios.
F15 confessed that \emph{``even someone in the weeds of this has a hard time knowing what they’re looking at with the map.''} 

\subsubsection{Tension Between the Individual and Collective}\label{finding_tension}

Three participants wanted clarity on how boundary changes would impact their own families, while five participants wanted to look at the data from a birds-eye view.
Regarding individual impacts, F11 wants \emph{``messages to be written in a way that makes it clear how information is pertinent to them and why.''}
F15 agreed, saying \emph{``people want information specific to them.''}
In contrast, F12 wanted access to data on how proposed boundaries would change community-level SES before sharing their feedback because \emph{``the whole point of public school is to give everyone an equal chance.''}
F14 emphasized how they would vote for a particular scenario because \emph{``it helps with the longevity of other schools''} even if they end up transferring their kids to private schools. 
Five participants directly touched on the tension between the individual and the collective.
F7 felt that everyone, including themselves, \emph{``will vote in their best interest but wants people to think back to what they agreed upon as a community.''}
F10 expressed a willingness to sacrifice individual benefits for the collective good: \emph{``I'm willing to take on the challenge of losing some [people in the middle of the SES spectrum] so that the relief high school gets some.''}

Seven participants expressed a desire to know what other people are saying about a boundary change.
F2 wanted to know what others said to \emph{``see how hard they will have to work to change the decision if the current majority isn’t what they want''} so they can build their own coalition. 
Less active participants, like F5 and F1, wanted to see other people's opinions to be \emph{``more aware of what's going on''} (F5) and for \emph{``the possibility of seeing a good idea from someone else that they would support''} (F1).
Two participants wanted increased transparency to better contextualize community feedback, such as understanding whether the people who voted for a scenario live in the impacted area.
F12 shared a \emph{``concern for the manipulation of survey responses''} and how it could be \emph{``misleading to see only percentages of how many people voted for what choice.''}

\subsubsection{Distrust Between the Community and the District}\label{finding_distrust}

Six participants shared concerns that they aren't being heard by the school district. 
F1 said that it \emph{``doesn’t always feel like [the school district] is using feedback.''}
F2 \emph{``doesn’t believe or trust that [the school district] reads the surveys.''}
F3 agreed: \emph{``Emails are being sent into a void.''}
F7 said the surveys are a \emph{``black box''} in that \emph{``you aren't sure how your responses impacted the decision.''} 
F6 said they lacked confidence in the school district and \emph{``won't bother to fill [surveys] out, which results in less equitable feedback.''} 
As a result of not feeling heard, parents inundate the School Board and Student Assignment Team with messages.
For example, the Director of the Student Assignment Team mentioned that a few parents emailed and texted her every day to share their feedback and ask for updates on the redistricting process.
To address the feeling of not being heard, participants wanted feedback mechanisms to be more \emph{``open-ended to give constructive feedback''} (F10).
Similarly, F4 said that the current \emph{``surveys don't allow for nuance''}, and F8 said they want to explain \emph{``why they chose [a particular] answer.''} 


These comments about not being heard by the district echo a broader lack of trust between the community and the school district.
Participants said the school district \emph{``does whatever the wealthy want''} (F8) and that \emph{``secret meetings with some people...builds mistrust''} (F7).
Several actions during the rezoning process contributed to the lack of trust. 
F3 recalled that the school district had \emph{``wrong data about what neighborhoods went to what school,''} which made the community \emph{``distrust the validity of [the district's] data.''}
F12 said the district \emph{``shot before they aimed, saying one thing then coming back to say another.''}
One factor that contributes to the lack of trust is not understanding why certain decisions are made.
Two participants suggested framing boundary changes in terms of the four Board priorities. 
F7 wants \emph{``survey questions to directly support the 4 pillars, starting with a question asking: `Do you agree with these pillars?'''}
F10 proposed creating a \emph{``rubric with weights for each pillar, which could help guide rezoning discussions and decisions.''}

\subsection{Design Questions}

Based on our formative study, we identified three design questions to guide our system design.
First, we observed a tension between the individual and the collective.
In turn, we explored ways to help community members understand how boundary changes impact their own families and the community at large, resulting in the following question: \textbf{How does an interface that depicts various school rezoning scenarios, with a focus on its impact on the district as a whole, prompt families in that district to consider the effect of their policy preferences on other families? (DQ-1)}
Almost half of the participants wanted the ability to view other people's feedback for various reasons.
Instead of giving the most active parents access to all the feedback, so they can promote their agendas, we explored more intentional ways of sharing different and unique perspectives to increase awareness and reflection.
This led to the following question: \textbf{When exposed to individual stories of families who are adversely affected by a policy priority, how do families who support those policies react to their role in creating a system that hurts another family? (DQ-2)}
In response to participants wanting to understand how a boundary change relates to the Board priorities, we explored: \textbf{If the district presents its rezoning goals with a shared vocabulary or set of community values, how do families perceive the district's intentions? (DQ-3)}
We delved into these three design questions while developing the BoundarEase platform.
In addition, we attempted to address some of the usability challenges in the original map interface and to devise a more nuanced way for people to share their feedback to feel more heard. 

\section{The BoundarEase System}
Keeping our design questions and user pain points in mind, we wanted to improve the community's experience of understanding and providing feedback on potential boundary changes; thus, we created a low-tech, mobile-friendly solution aiming to make those actions more accessible. A web-based solution better enables interactivity, personalized information, asynchronous participation, and translation into other languages. Furthermore, the information design was carefully considered to be scalable to other districts that may not have the resources to implement a new sociotechnical system on their own.  While technology can sometimes exacerbate inequities in access to information, it can also help automate or otherwise reduce burdens for manual analyses to make sense of how certain policies might impact families.  These are analyses that more resourced members of the community may be able to conduct---and in fact, did conduct in the district---but that others may not have the time or expertise to engage in.  In this way, our hypothesis was that a technology-enabled platform can serve as a useful scaffold for making sense of complex processes like student assignment policymaking---and in turn, help mitigate inequities in access to information and understanding in this policy-making process.  Still, a technology-enabled system may also have unintended consequences that threaten our equity objectives; we return to these possibilities in the Discussion section. 

At the foundation of BoundarEase are the Board's four policy pillars: socioeconomic (SES) diversity, home-to-school distance, feeder patterns, and utilization. Definitions for the pillars can also be found in Table~\ref{tab:pillardefinitions} of the Appendix. For the purpose of this study, BoundarEase asks parents how they prioritize each of the four pillars and then shows them algorithmically generated attendance boundaries that reflect their preferences.  To do this, we used the redistricting algorithms from~\cite{gillani2023redrawing} as a starting point.  The original algorithm's objective function is modified to consider the user's preferred priority rank for each pillar. We apply the Softmax function to the vector of ranks per pillar, placing significantly higher weight on the pillars that families rank as their top two versus bottom two. The top-ranked pillar is assigned a weight of 0.64, the second rank 0.24, the third 0.09, and the lowest ranked 0.03. The weights sum to 1.0.


\subsection{System Description}
We implemented the front-end of the system using various JavaScript libraries, including React~\cite{react}, Material UI (MUI)~\cite{mui}, D3.js~\cite{d3js}, and React Leaflet~\cite{reactleaflet}, and communicated with the back-end database using Flask~\cite{flask} and SQLAlchemy~\cite{sqlalchemy}. 
Unfortunately, the signed data agreement with the school district came following the conclusion of the project.  Therefore, datasets for school populations, school attendance boundaries, and travel times were drawn from both public and purchased datasets described in~\cite{gillani2023redrawing}.

\subsubsection{Start Page}
\label{sec:startpage}
The initial page (see Fig. \ref{fig:startpage} in the Appendix) asks users for two inputs: how they rank the four pillars---which they input by dragging cards into a desired order--- and their address. This information allows BoundarEase to provide customized statistics, streamlining the users' understanding of how they would be affected (Finding \ref{finding_tension}) in a boundary proposal that is optimized based on their policy preferences.

\begin{figure}[h]
\centering
\includegraphics[width=0.8\linewidth]{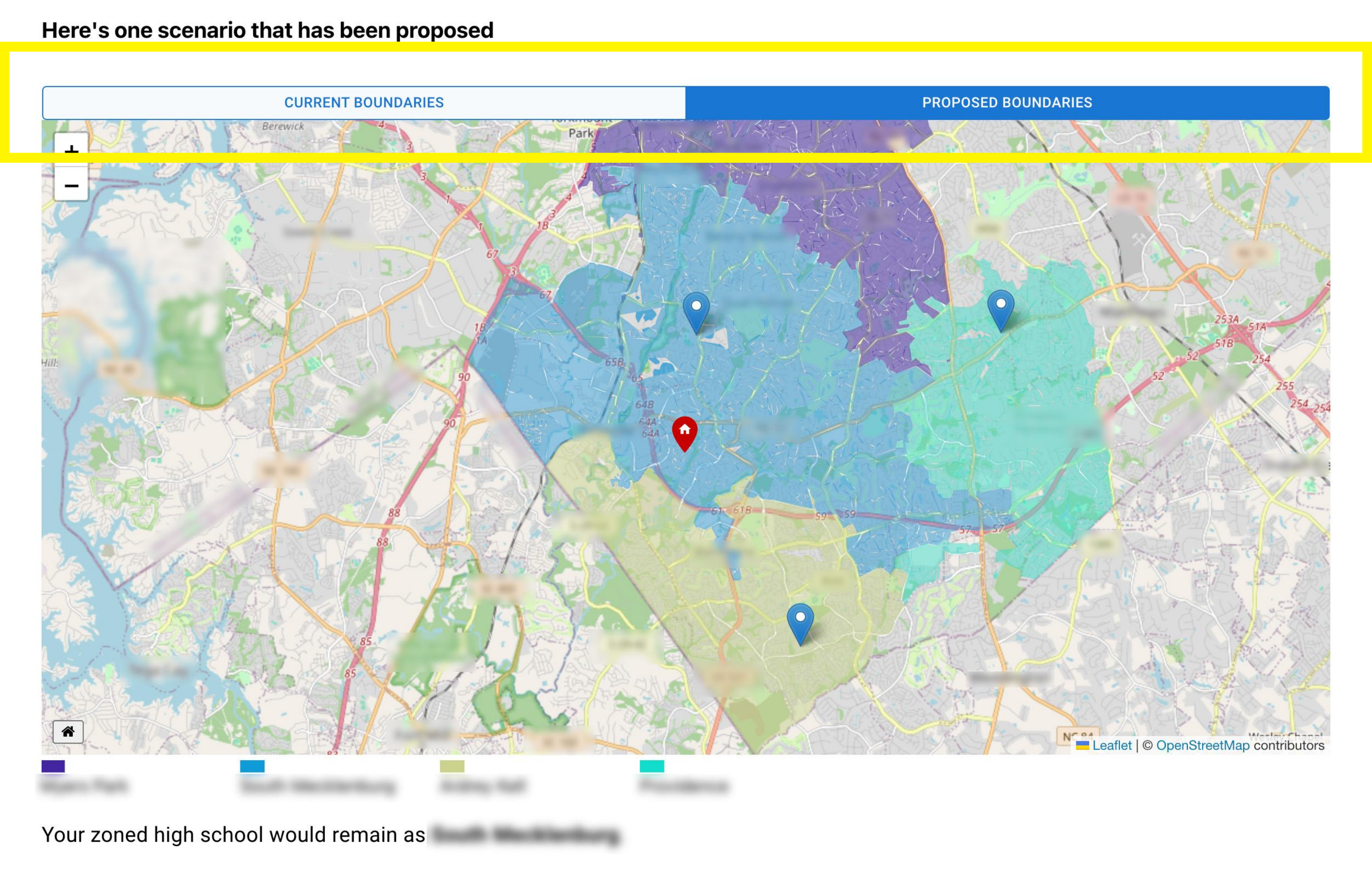}
\caption{The interactive map. Users can toggle between the current and proposed boundaries.}
\label{fig:interactivemap}
\Description{Screenshot containing interactive map. Two toggle buttons span the width of the map. Within the map are the boundaries, location markers, grouped ''plus'' and ''minus'' buttons to zoom in and out, and a recentering button. Beneath is a legend of colored rectangles and the corresponding school for that boundary.}
\end{figure}


\subsubsection{Interactive Map}
\label{sec:map}
After users enter their information, they are shown a set of attendance boundaries based on their pillar ranking. At the top of the page is an interactive map (Fig.~\ref{fig:interactivemap}) depicting the boundaries, distinguished through semi-transparent colors, and markers highlighting the impacted schools' locations and the user's address. Users can toggle between the current and proposed boundaries, as well as zoom in on streets and individual buildings. This simple functionality was motivated by the pain points described in Finding~\ref{finding_usability}, in which formative study participants shared that the overlays--- and even the map itself--- were often illegible and ineffective. 

\begin{figure}[h]
    \centering
    \includegraphics[width=0.7\linewidth]{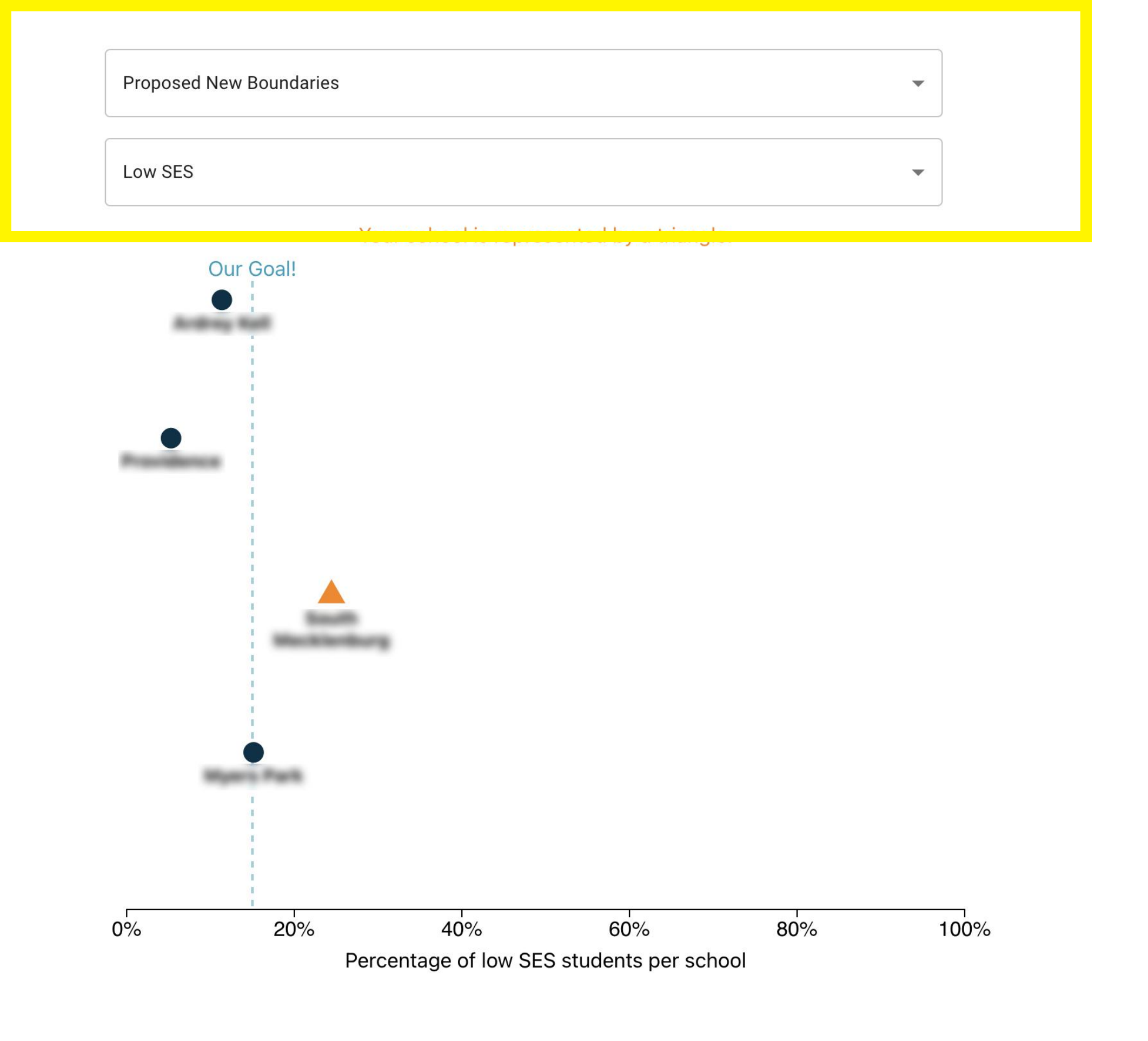}
    \caption{Visualization showing the percentage of students within each socioeconomic category across impacted schools.}
    \label{fig:ses_viz}
    \Description{The SES diagram has two vertically-placed dropdowns. The first controls whether the data being shown is for the current or proposed new boundaries. The second controls which SES is being shown, low, mid, or high. At the bottom of the visualization is a number line with 20\% intervals, 0\% at the left-most end and 100\% at the right-most end, and a label ''Percentage of low SES students per school''. Perpendicular to the number line is a dashed line labeled ''Our Goal''. Black circles labeled with school names are placed above the number line with respect to the percentage of students in that SES at that school. The user's school is an orange triangle.}
\end{figure}

\begin{figure}[h]
    \includegraphics[width=0.7\linewidth]{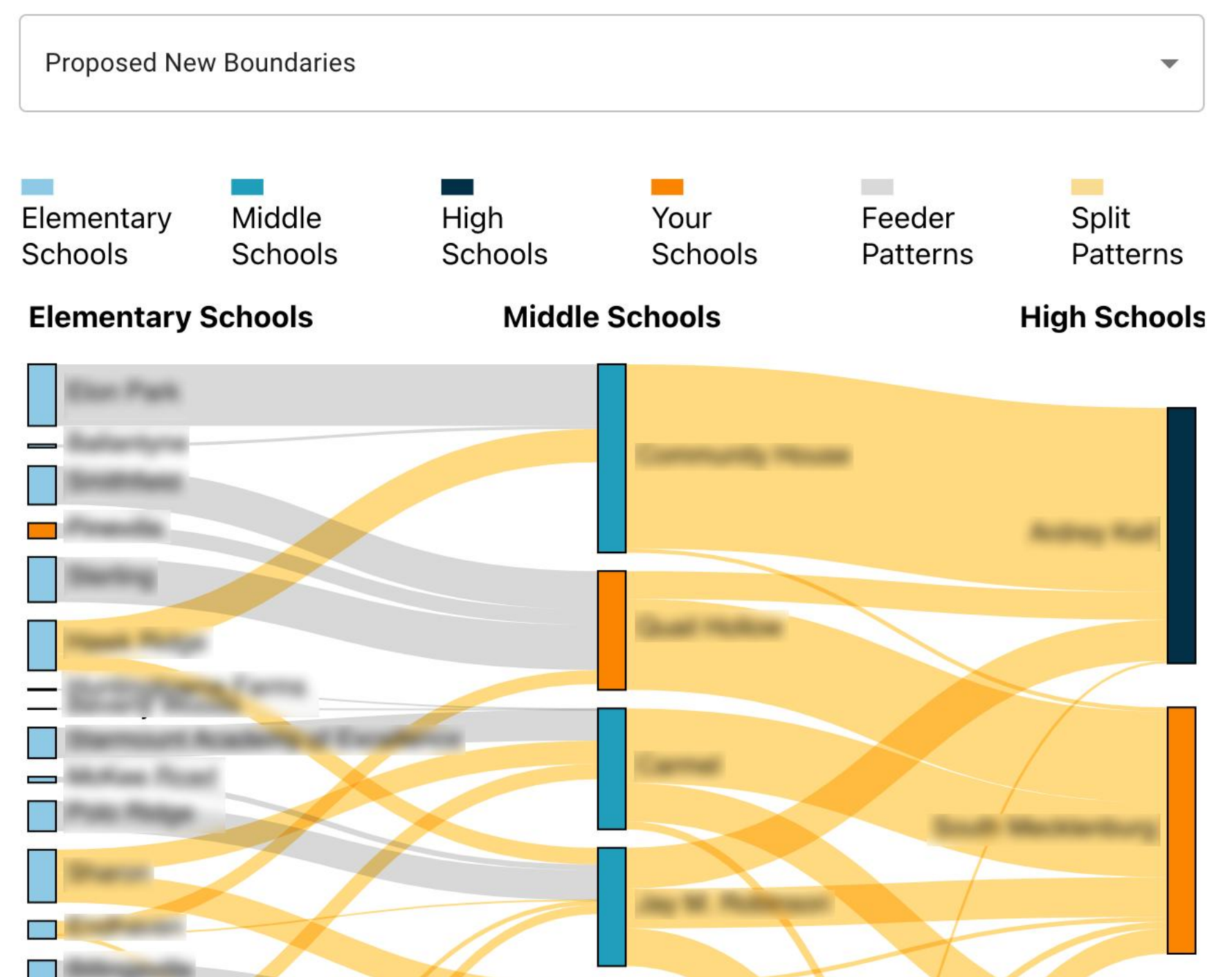}
    \caption{Sankey diagram visualizing feeder patterns across the impacted schools. Hovering over a pathway will reveal the number of students going from one school to the next.}
    \label{fig:feeder_viz}
    \Description{The Feeder visual has one dropdown to toggle between the current and proposed new boundaries. Beneath the dropdown is a legend of colored rectangles labeled ''elementary,'' ''middle,'' '' high,'' or ''your'' schools, ''feeder patterns,'' and ''split patterns''. Then there are three columns of nodes. Each column corresponds to a level of school and each node within the column represents a school and is labelled accordingly.}
\end{figure}

\begin{figure}[h]
    \includegraphics[width=0.7\linewidth]{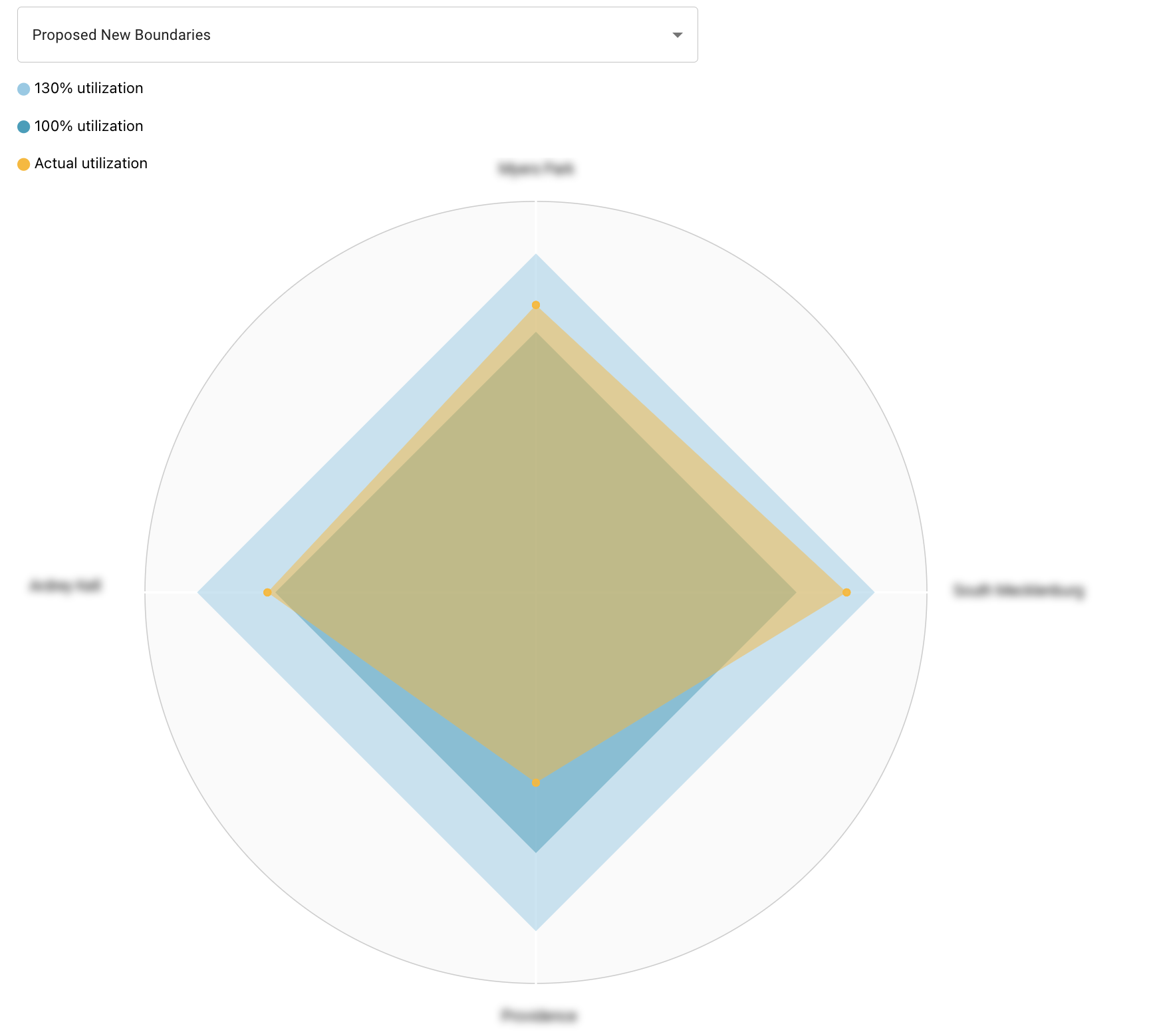}
    \caption{Radar chart depicting the enrollment and capacity at all the impacted schools.}
    \label{fig:utilization_viz}
    \Description{The Utilization chart has one dropdown to toggle between current and proposed new boundaries. It has a legend of colored circles, labeled ''100\% utilization'', ''130\% utilization'' and ''actual utilization''. Beneath the legend is the radar chart with an impacted school name at the top, bottom, left, and right. Within the chart are three semi-transparent polygons, a light blue representing the 130\% utilization mark, a darker blue representing the 100\% mark, and a yellow representing the actual utilizations.}
\end{figure}

\subsubsection{Communication of Policy Impact}
\label{sec:policyimpact}
Following the map, BoundarEase frames the impact of the boundaries through the four pillars. To streamline this information, we give each pillar its own section which follows a standardized format: the pillar name, a short definition of the pillar, and data describing the expected impacts of the new boundaries at both a personal and district-wide level. The data is conveyed through concise, dynamic text, and bolded keywords/numbers (see Section \ref{appendix:system} in the Appendix for each pillar's text). Some sections include additional information about why a pillar may be valuable to consider when redrawing attendance boundaries. We also developed accompanying interactive visualizations (shown in Fig.~\ref{fig:ses_viz}, ~\ref{fig:feeder_viz}, and ~\ref{fig:utilization_viz}) for three of the four pillars, exploring different approaches to investigate their potential effectiveness.

\paragraph{SES Diversity}

The visualization in Fig.~\ref{fig:ses_viz} shows how the proposed change affects SES distribution in the affected schools. Users can change the dropdown value to switch between the current boundaries and proposed new boundaries, as well as select which SES level to visualize, starting with ``Low SES''. This allows the user to see how close the selected SES of a school is to the SES distribution of the region being rezoned---a notion of ``evenness'' that is similar to many existing definitions of segregation, like the popular dissimilarity index~\cite{massey1988dimensions}. Hovering over a school shows the full SES distribution for that school.


\paragraph{Home-To-School Distance}

The home-to-school distance pillar is kept as text only. While it does not account for traffic patterns (similar to the analyses in~\cite{gillani2023redrawing} due to data limitations), we opted to use travel time instead of distance as it felt like a more relevant metric for users (based on feedback from the formative study). The system shows the user their current travel time and what their new estimated travel time would be under the proposed boundaries, which we calculate using a method from previous work ~\cite{gillani2023redrawing}. We also share the number and percentage of students in the district who would experience an increase and decrease in travel times, respectively.

\paragraph{Feeder Patterns}

As shown in Fig.~\ref{fig:feeder_viz}, we use a Sankey diagram to visualize feeder patterns across the impacted schools (i.e. which high and middle schools students from each middle and elementary school are expected to transition to, respectively), with users being able to toggle between the current and proposed boundaries. Going from left to right, the nodes of the Sankey represent the elementary, middle, and high school levels, with orange nodes highlighting the user's feeder pattern. Gray is the default color of the pathways between the nodes, while pale yellow pathways indicate a split. Hovering over a pathway will reveal the number of students expected to transition from one school to the next. Though the Sankey is a less common diagram, we felt it would be the most intuitive way to visualize student flows between schools.





\paragraph{Utilization}
The visualization depicted in Fig.~\ref{fig:utilization_viz} is a radar chart. Each vertex is an impacted school, and the three shapes show the benchmarks as defined by the district: the 130\% utilization limit (light blue), the 100\% utilization limit (dark blue), and the actual/projected utilizations (yellow).  Users can toggle between the utilizations of the current and proposed boundaries, and they can hover over the radar to see the utilization percentages and student enrollment. Schools may allow more than 100\% utilization if they can accommodate additional students through mobile classrooms. This was a more speculative visual, as we wanted to evaluate if community members could interpret less common visualization methods.





\subsubsection{Feedback Collection}
\label{sec:feedbackcollection}
\begin{figure}[h]
\includegraphics[width=0.7\linewidth]{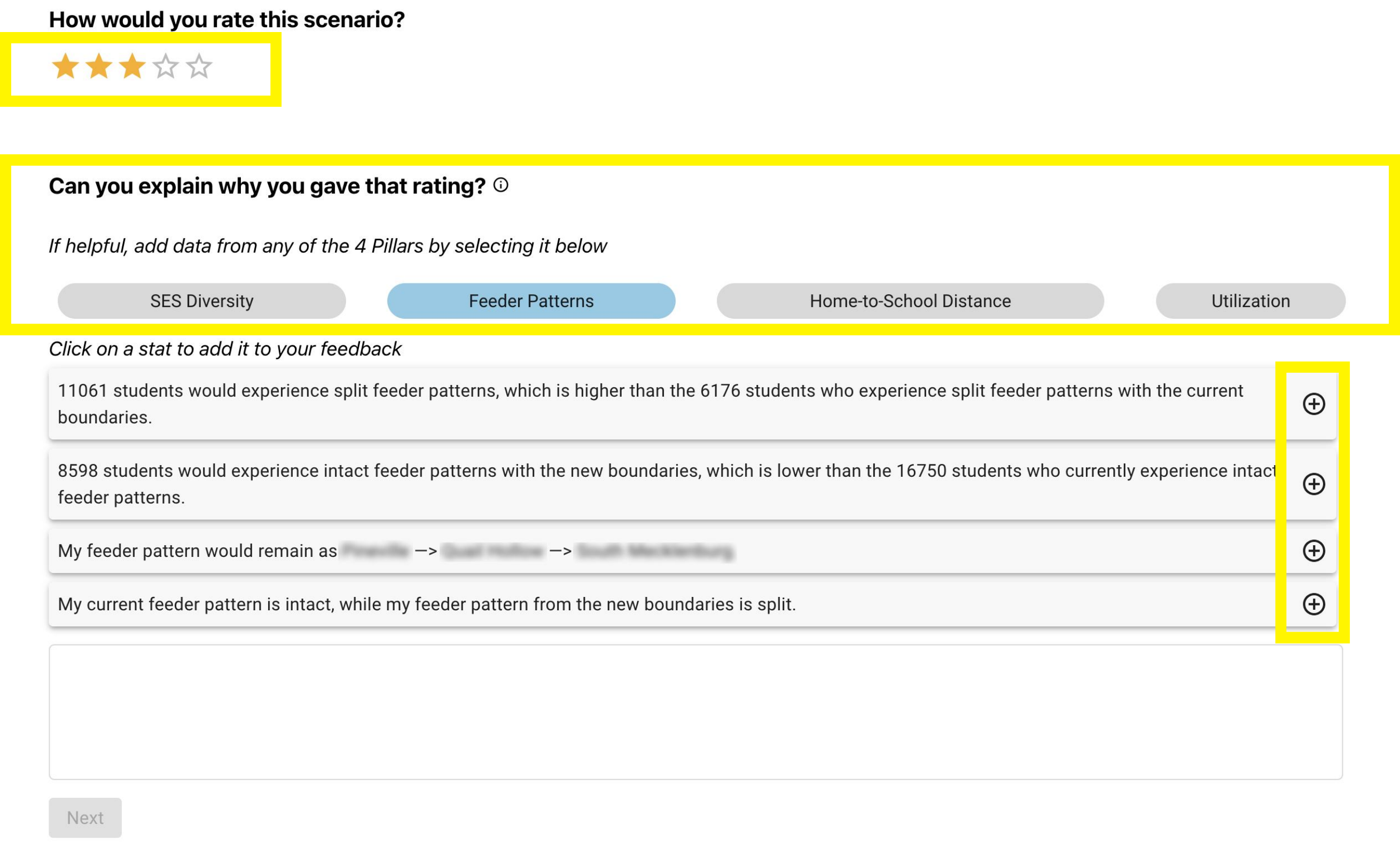} 
\caption{The interface for collecting feedback. Users are asked to rate the proposed boundaries on a 5-star scale after viewing the data on the four pillars. A multiline text field enables them to freely type, with the option to add statistics they find relevant.}
\label{fig:feedback_collection}
\Description{The interface for collecting feedback. At the top is the prompt, ''How would you rate this scenario?'' and outlines of five stars underneath. Below those is the text, ''Can you explain why you gave that rating?'' In a smaller font beneath that is the sentence, ''If helpful, add data from any of the 4 Pillars by selecting it below.'' There is then a row of gray buttons, each labeled with a pillar. One is selected, as indicated in light blue, to reveal the text ''Click on a stat to add to your feedback,'' followed by the statistics for that pillar. Each statistic is on a card with a plus sign at the right end. Below the statistics is the text field and a disabled Next button.}
\end{figure}

As shown in Fig.~\ref{fig:feedback_collection}, users are asked to rate the proposed boundaries on a five-star scale after viewing the data on the four pillars. Once they do so, they can give open-ended feedback in a text field (inspired by findings in Section \ref{finding_distrust}). 
We also provide the statistics that they read earlier in the interface, which they can click to add to the feedback text field and edit as desired.
%
%
%
%
%
By structuring feedback collection this way, we hoped to capture the richness of qualitative, open-ended feedback while still grounding this feedback in data, adding healthy frictions that may impede purely emotional ---and perhaps, less constructive---input. Conversely, for users who do not have a strong opinion, the same structure may help scaffold their thoughts.

\begin{figure}[h]
\includegraphics[width=0.7\linewidth]{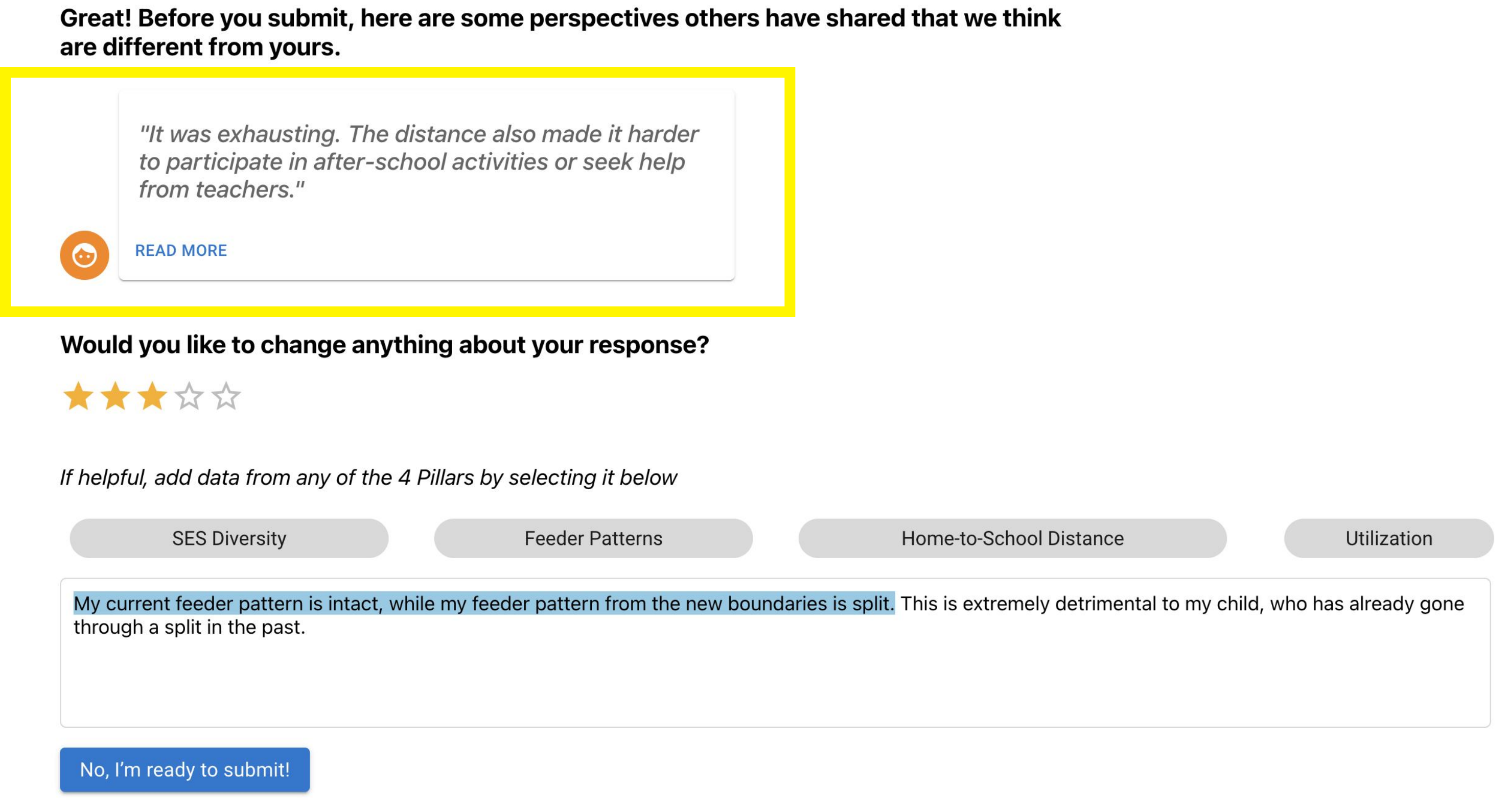} 
\caption{Users are shown an expandable snippet of someone else's perspective, which we state may be vastly different from the user’s opinion, based on their prioritization of the pillars that was gathered on the first page. If desired, users are able to change their original rating and the feedback via the same interface structure before submitting.}
\label{fig:perspective_getting}
\Description{The final page. At the top is the text, ''Great! Before you submit, here are some perspectives others have shared that we think are different from yours.'' Beneath that is the perspective. It includes a short quote on a card. At the bottom of the card is a Read More button. To the left of and bottom-aligned with the card is an icon of a face silhouette placed on a circle. Below this is the same interface for feedback collection, aside from some differences in text. Above the star rating the text instead is ''Would you like to change anything about your response?'' and the button at the bottom says ''No, I'm ready to submit!''}
\end{figure}

\subsubsection{Exposure to Other Perspectives}
\label{sec:thirdpage}
After entering their feedback, users are brought to the final page depicted in Fig.~\ref{fig:perspective_getting}. They are shown an expandable snippet of someone else's perspective, which we state may be different from the user's opinion based on their prioritization of the pillars, which was gathered on the first page. For the purpose of this study, we include only one perspective---a student persona's testimonial generated by GPT 3.5---to obtain a sense of how a hypothetical student with different priorities over the pillars might respond. The hypothetical perspective is meant to illustrate how deprioritizing certain pillars might yield adverse impacts for other students/families who have different priorities. In later iterations, we envision replacing these synthetic perspectives with actual input from students and parents. This feature was informed by prior perspective-getting research, which has shown that hearing narratives from/about out-group members can help reduce prejudice and exclusionary attitudes~\cite{kalla2023narrative}.
If desired, users can change their original rating and their feedback before submitting. 




\section{User Study}

We evaluated the use of BoundarEase to collect feedback on possible boundary changes. 
We paid particular attention to how BoundarEase addresses our design questions of emphasizing the connection between the individual and the collective (DQ-1), incorporating perspective-getting (DQ-2), and framing changes in terms of the Board priorities (DQ-3).

\subsection{User Study Method}

We conducted semi-structured interviews with 12 parents who participated in the redistricting that we studied during the formative study.
We decided to conduct interviews instead of running an experiment or field study because we wanted to speak with people who recently went through a boundary change to get their perspective on how the previous engagement process compares with BoundarEase. 
We provided \$30 Amazon gift cards to participants for an hour-long conversation.
The interview protocol received Institutional Review Board approval.
We reached out to a total of 34 parents through the Student Assignment Team;
12 of them agreed to give feedback on the platform.
We refer to the participants from P1 to P12. 
Six of these participants also gave feedback during the formative study. 
Nine participants were active during the rezoning process, while three participants said they were moderately or less involved. 
All except one participant is female; eight participants lived in higher-SES areas, while four lived in lower-SES ones. 
Eight participants also had children attending multiple schools with grades ranging from preschool to 12th grade. 
Additional participant details are in Table~\ref{tab:userparticipants} in the Appendix.

At the beginning of the interview, participants were asked to rank the four Board priorities (feeder patterns, home-to-school distance, SES diversity, and utilization) by order of importance. 
The participants held a diverse range of opinions on their policy priorities, as illustrated in Table~\ref{tab:users}.
One group of participants valued home-to-school distance the most, while another group valued SES diversity the most.

\begin{table}
  \caption{Participant rankings for the four Board priorities that guide student assignment policy changes. The numbers are counts for how many participants ordered each priority in a certain way. First place corresponds to the priority that participants found most important. }
  \label{tab:users}
  \begin{tabular}{l|cccc}
    \toprule
    \textbf{Ranking}&\textbf{Feeder patterns}&\textbf{Home-to-school distance}&\textbf{SES diversity}&\textbf{Utilization}\\
    \midrule
    First place&2&5&5&0\\
    Second place&8&2&1&1\\
    Third place&2&2&3&5\\
    Fourth place&0&3&3&6\\
  \bottomrule
\end{tabular}
\end{table}

After providing context on the project, we gave participants access to the BoundarEase platform.
Participants were prompted to think aloud and interact with the first two pages of the platform (Sections~\ref{sec:startpage} to \ref{sec:feedbackcollection}). 
From there, we asked how the experience of seeing potential boundary changes and providing feedback on them compared to what the school district recently did.
After they obtained an initial impression of BoundarEase, we asked participants to think aloud and interact with the third page of the platform (Section~\ref{sec:thirdpage}).
After arriving at a simple ``thank you'' page, we verbally asked a series of Likert-scale questions to compare the platform with the district's status quo.
Before ending the interview, we shared information about how the new boundaries were algorithmically generated and asked for feedback on this approach of generating possible boundary changes.
We also asked for hopes and concerns about using BoundarEase for community engagement.  
See supplementary materials for the questions we asked.

Interviews were conducted remotely over video calls, which were recorded and transcribed using an auto-transcription service.
We also took extensive notes during the conversations.
We analyzed the interviews using thematic analysis~\cite{braun2006using}.
First, we segmented the interview transcripts into paragraphs, which formed our units of analysis.
Then, each member of the research team independently read all of the interview transcripts and notes and wrote down possible topics in Figma, an online collaborative whiteboarding space~\cite{figma}.
The research team met synchronously to cluster the initial topics to form categories, which we refined into a two-level hierarchical codebook consisting of 15 parent codes and 38 child codes. 
Parent codes represent general topics (e.g. ease of use and data vs human), and child codes (e.g. accessibility and data-driven) represent various facets of a topic. 
See supplementary materials for our codebook. 
From there, the first two authors collaboratively applied the codebook to three interviews to refine the definitions and address possible ambiguities.
After updating the codebook, the first two authors independently coded the remaining nine interviews using a spreadsheet application.
All disagreements were resolved through periodic discussions during the coding process.
The final inter-coder reliability was 0.68, measured using Cohen's Kappa~\cite{cohen1960coefficient} and indicating strong agreement.
The entire research team had several discussions to condense the codebook into the themes presented in Table~\ref{tab:themes}.  

\subsection{Results}

Our analysis of the interview transcripts and notes reveals key insights into how BoundarEase could enable community members to understand and share their views about a prospective attendance boundary change. 
Table~\ref{tab:themes} summarizes the main themes that we identified from thematic analysis.
We expand on these themes in the following sections while discussing opportunities for improvement. 

\begin{table*}
  \caption{Table depicting the themes that emerged from our analysis. A few of the larger themes include more specific sub-themes. Each row also includes a description and a list of codes associated with a theme or sub-theme. Parent codes are denoted with an asterisk.}
  \label{tab:themes}
  \begin{tabular}{@{}p{0.13\linewidth}|p{0.13\linewidth}p{0.383\linewidth}p{0.33\linewidth}}
    \toprule
    \textbf{Themes} & \textbf{Sub-Themes} & \textbf{Codes} & \textbf{Descriptions}\\
    \midrule
    Informed decision-making & General awareness & Informed decision-making*, lack of awareness* & Users' level of awareness about possible boundary changes\\ \cline{2-4}
    & Individual vs collective & Learning impact on community, learning impact on self, impressions on text descriptions, impressions on visualizations & Learning the impact of a boundary change on oneself and the community, as well as drawing connections between the two\\ \cline{2-4}
    & Perspective-getting & Impressions on the perspective-getting feature, impressions on the modifying feedback feature & Reactions to being exposed to perspectives different from one’s own\\ \cline{2-4}
    & Shared \newline vocabulary & Data-driven & Reactions to framing a boundary change in terms of a predefined set of community values\\ \hline
    Sharing feedback on boundary changes & --- & Sharing feedback, scaffolding, data-driven, impressions on adding stats feature, impressions on rating feature & The ability to share preferences, feedback, and concerns on potential boundary changes\\ \hline
    Concerns on platform’s role in \newline process & General \newline concerns & Poor communication, transparency, community manipulation, school district manipulation, collective reactions to process*, negative impressions on the algorithm, human knowledge & Concerns about integrating the platform into the community engagement process for boundary planning\\ \cline{2-4}
    & Process \newline structure & Structure of process*, process length & Recommendations on how to structure the community engagement process\\ \hline
    Interface usability and accessibility & Current strengths & Conciseness, understandability, accessibility, task time, responsiveness*, positive feature impressions* & Ability to efficiently understand and interact with platform features \\ \cline{2-4}
    & Improvement suggestions & Specific suggestions*, points of confusion*, bugs*, negative feature impressions* & Suggestions to improve the platform\\
    \bottomrule
  \end{tabular}
\end{table*}

\subsubsection{Enhancing Informed Decision-Making}

\paragraph{Addressing Varying Levels of Awareness}

Participants had varying levels of awareness about possible boundary changes.
Nine participants were heavily involved in the rezoning process, saying that their engagement levels were essentially \emph{``second jobs.''}
Three participants were less active during the process.
P2 observed a general lack of awareness from many parents with whom she interacted, sharing that \emph{``people had no idea how the rezoning was impacting them. Even now, people don’t know what happened.''}
An overload of information from the school district was one factor that discouraged less active parents from being fully aware of possible boundary changes.
P1 said that she relied on the parent-teacher association (PTA) to stay in the loop: \emph{``We have a very active PTA, and our PTA President was going to these meetings and bringing information back to us along with another parent group because I don't have time to do that. I work full-time. If it would not have been for that information, I would not have known half of what was going on.''}




Four participants expressed that the platform would help the greater community become more knowledgeable on boundary changes and feel more empowered to give feedback.
P8 said the platform \emph{``would help people understand what's going on a little bit better.''}
P9 agrees: \emph{``In just a few minutes, I could be informed of the whole thing. I could visually have an idea of what they were trying to communicate. There were a lot of questions that are very well explained here. It was clearer than the meetings.''}
P3 felt hopeful about the school district's interest in BoundarEase since the platform is \emph{``giving you the information. [The district] wants you to understand it and wants to get your feedback on each part of it.''} 
P1 was \emph{``happy that the school system is even considering a platform like this, which I think is a lot more transparent, provides a lot more information for parents to be able to make informed decisions and give informed feedback.''}

\paragraph{Balancing Personal Interests and Community Impact}

From the formative study, we observed a tension between the individual and the collective. 
While creating BoundarEase, we attempted to explore this tension through two design questions (DQ-1 and DQ-2).
We created interactive visualizations to help families see the broader community-level impacts of a possible boundary change (DQ-1).
The perspective-getting feature was another way of exposing users to other perspectives (DQ-2). 
12 out of 12 participants thought that the clarity of how new boundaries would affect their families and others was better than the status quo.
In addition, 82\% of participants agreed that BoundarEase helped them understand how others might be affected by a rezoning and/or what they care about.

The platform effectively explains how a boundary change impacts an individual user. 
P12 observed that \emph{``it's really summarizing my data. It was really clear with what my address did.''}
P2 compares BoundarEase's ability to explain individual impacts with the status quo: \emph{``The fact that you could put in your address, and then immediately came up as your zoned high school would remain as [this school] would have been helpful... I probably had 100 to 120 text messages or Facebook messages from people I have never met saying, `I don't know how to use the interactive map. Can you just put in my address and tell me where I'm zoned?'''}


A majority of participants mentioned the ability to learn about community-level impacts through the text descriptions and visualizations under each pillar.
The feeder pattern visualization was especially effective in showing the impact of a boundary change on the collective.
P10 said, \emph{``The feeder pattern graph really showed how schools are getting chopped up. Being able to show all the data in one place would give everyone a broad picture of what happens.''}
P11 more broadly reflected on the importance of showing the data for multiple schools: \emph{``I think that's helpful because even if you're kind of focused on one area because of your family, obviously you want to try to help the community as a whole and you want to make sure that all of the schools are as equitable as possible.''} 
%

P4 wanted to test out different rankings of the four pillars to re-think her perspective: \emph{``That's also helping you see what other people are seeing and it may help you to rethink how you're coming about it.''} 
P3 mentioned the importance of providing perspective to reduce selfishness: \emph{``What we saw during this whole process was a lot of people focusing on themselves. 
A lot of times people look at themselves, saying, `Oh, well, I have an increase', but in reality, you're part of 1\% that has an increase. It gives perspective.''} 
P1 emphasized how showing the collective impact, particularly on those adversely affected by a proposal, could make people more empathetic and less emotional about a boundary change.

\begin{displayquote}
\emph{``When you don't see a negative impact on other people, it's very easy to be selfish and just advocate for what's best for you or for your student, but you should take pause. If a choice that you're making for your student is gonna negatively impact a lot of other students, I think we all have a responsibility to our communities, not just to ourselves. I definitely think, having that comparative information is very important, and I hope it would help people be a little bit more practical and rational because these types of topics can get very emotionally charged.''}
\end{displayquote}

Participants also noted several limitations in how the community-level impacts are explained in the platform.
P7 mentioned that the summaries and visualizations were not enough for her to understand \emph{``how somebody with another high school would be changed.''}
P5 wanted to see a more detailed breakdown of SES levels for every school.
P1 wanted to see data representing the entire school district instead of just the impacted region: \emph{``Sometimes it helps to be able to zoom out a little bit and look at what it's like across our entire district to see if we're in line with what's happening across the district, and not to feel like we're being singled out.''}


\paragraph{Reactions When Exposed to Different Perspectives}

We observed a wide range of reactions towards the perspective-getting feature (DQ-2). 
Eight participants appreciated the ability to view perspectives different from their own.
P11 wanted to see other perspectives to fill in possible knowledge gaps since she \emph{``might be missing something, or missing a question, or missing a thought.''}
P8 thought the feature could make some people less siloed: \emph{``It's nice to have [other perspectives] cause if you're not going to in-person community engagement sessions or reading feedback online or engaging in different sessions, then you don't ever have a chance.''}
%
%
P9 said that it is helpful \emph{``to have someone else's experiences here to keep all those possibilities in mind, not only your own circumstances.''} 
P2 recalled the previous rezoning process getting \emph{``really ugly''} and mentioned that seeing other people's stories \emph{``in print and in a nice way while you [are] taking the survey, especially at the beginning, before it got so emotional, would have been really helpful.''}
P4 suggests that the feature may \emph{``make people more willing to give concessions on the [scenarios] that they scored lower.''} 

Though the perspective-getting feature may be appreciated by some, P8 predicts that \emph{``other people will just say, this is my perspective and submit and not think about it.''}
P5 is less optimistic about how the feature would be used in reality.

\begin{displayquote}
\emph{``I think the reality is, people are not gonna see the alternate perspective and say, `Oh, that's really interesting. I didn't think about that.' It's just such a heated thing, and people for the most part want what's best for them and their family and their school.''}
\end{displayquote}
%
P7 shared another concern: \emph{``People who are more impressionable might end up being influenced.''} 
%
To alleviate some of these concerns, two participants recommended adding more than one perspective and creating more positive stories.
P11 suggested displaying stories with similar and different perspectives to \emph{``get a big picture of what others are thinking.''}
P7 felt that having a \emph{``quote promoting each of these [pillars] would feel a lot less leading.''} 

\paragraph{Aligning Boundary Change Proposals with Shared Policy Goals}
One of our design questions involved framing a boundary change in terms of the four pillars that guide the school district's student assignment policies (DQ-3).
11 out of 12 participants thought that the clarity of how new boundaries align with the four pillars was better than the status quo. 
P3 elaborated on the importance of having a shared vocabulary: \emph{``All of the numbers and the data is available for you in a digestible form across the four pillars, so if this is going to be [the school district's] drive in any of the decisions they're making, it's nice to be able to see how they're meeting any of [the pillars].''}
%
%
However, framing a possible boundary change with respect to the four pillars is not sufficient in effectively communicating shared values. 
P12 was concerned about a lack of criteria for each pillar.
She recalls that \emph{``there became a huge ordeal with the SES, and no one had a set criteria that they said they were actually following.''}
P4 brought up concerns about constantly redefining the criteria during the district's rezoning process: \emph{``They kept moving where the line was for utilization like they would give one number, and then they would give another number. You can make the data say whatever you want the data to say.''}
P10 wanted to go beyond the four pillars and have the platform more clearly explain how the pillars shaped the algorithmically generated boundaries: \emph{``I know it’s data, but it’s also children, so the parents need to understand how the determination was made.''} 

\subsubsection{Improved Feedback-Sharing Experience}

BoundarEase empowers users to give more specific, standardized feedback.
P3 talked about their experience at a recent community feedback session, where the lack of structure left parents confused about what approach they should take when sharing their thoughts: \emph{``Everyone focuses on saying why they don't like a certain proposal. And we were battling with, `Okay, well, we actually like this proposal, so do we give the same attention to the things we like in this? Or do we only give attention if we don't like something in it?' And then it was almost like we felt like we didn't have a voice.''} 
%
%
P2 highlighted how, unlike BoundarEase with its five-star scale, the previous process required binary answers: \emph{``I would much prefer to have done it this way than all the millions of surveys... I mean, really, the only choices we had in what we just went through was, `Do you agree with this scenario, or do you not agree with this scenario?'''}  
%

We also saw a strong appreciation for the ability to select specific data points to add to the feedback textbox (Fig.~\ref{fig:feedback_collection}). P1 stated: \emph{``I do think sometimes people have a hard time articulating their thoughts. So I think it's nice to provide some of these conclusive things based on what you know is happening.''} P5 echoed this sentiment, viewing the feature as \emph{``built-in talking points, [which] is a helpful way for the planning team to be able to quantify more where people are coming from.''}
P11 emphasized the convenience: \emph{``Some people just don't want to type a whole lot...[You could] drop in some of the context that you find to be beneficial without having to type it in.''} Along a similar vein, P9 mentioned that the number of statistics is ideal, saying, \emph{``it helps when one of your concerns is related to [one of the pillars]...But you don't have like 300 options here to choose.''}
P2 felt that the feature was successful in making feedback more constructive. She shared that \emph{``after reading 2,200 PDF pages of comments, there was an awful lot of emotion in those comments, and not a lot of data.''}

Three participants suggested breaking down feedback collection even further, such as by pillar or sentiment. P1 proposed \emph{``asking the participant to analyze a section of information at a time and give feedback on that section''}.
%
Others were thinking of how to make it more effective for both those analyzing feedback and those giving feedback. P3 focused on the former, saying, \emph{``If I look at this as if I'm working for [the school] and trying to gather data for this, the more words that are in my actual response, the harder it is to actually make sense of it or to be able to kinda organize it.''} She felt having a textbox for each pillar or keeping the statistics separate from the free text
could accomplish this. P8 wondered if splitting the feedback into pros and cons could be beneficial.

\subsubsection{Concerns on Platform's Role in Process}

\paragraph{Trust and Transparency Concerns}

Participants expressed a few concerns about integrating BoundarEase into the community engagement process.
A few of the concerns relate to the back-end algorithm, which generates boundary scenarios based on rankings of the four pillars.
P11 wondered how the algorithm would deal with multiple inputs that may be conflicting: \emph{``So many different people prioritize so many different things. I wonder how we could get to an actual boundary scenario that fits all the pillars.''} 
P8 noted that \emph{``the data can't tell the whole story''}, and she hopes \emph{``there are some people on the grounds that know the area that contribute to the maps.''} 
P11 wants the school district to fine-tune the algorithmically-generated maps since \emph{``[the algorithm] doesn't take into account the buses and the transportation issues that we have in our area.''}

Several concerns evolve around a general lack of trust in the school district. 
As P7 points out, \emph{``I feel like [the school district] pretends that they care and want to listen, and they just do their own thing.''}
To help mitigate distrust, participants touched upon the need for increased transparency.
P11 wanted to see the next steps displayed on the final thank you page: \emph{``People felt a little like it was a black box like I'm inputting my heart and soul into this, and then I don't know what's happening next. It would be cool to say something like, `The next step is we're compiling all of this data, and we'll be coming back to the community in X to X weeks.'''}
%

Another way to increase transparency is by sharing other people's feedback. 
P2 said, \emph{``I would want to know how many people rated this scenario five stars.''}
Signs of distrust were not only directed at the school district but also between community members.
P7 was worried that a map would be \emph{``generated based on very vocal parents.''}
P5 mentioned that during the previous engagement process \emph{``certain communities were just submitting over and over to skew the data.''}

\paragraph{Considerations Around Structure of Engagement Process}


Participants wanted to use the BoundarEase platform \emph{``early in the process before it gets really bad''} (P2). 
A few parents expressed concern about the possibility of people looking at different maps based on how they ranked the four pillars.
P1 said, \emph{``We all need to be looking at the same information presented the same way if we're really going to be weighing in and giving input on a change.''} 
P3 suggested a two-phase approach to structuring the community engagement process: \emph{``If there was a behind-the-scene algorithm where you took some type of survey prior to creating the proposal and say, `We took a survey of 500 people and 80\% cared about feeder patterns, and that's how this proposal was created.' And then that's what is given to everybody to vote on. I think that makes a little more sense than each individual person getting a different proposal to review and vote on.''}
%

With any multi-phase approach, engagement fatigue becomes a critical obstacle in reaching a wider group, particularly less active parents like P1, who said: \emph{``We get fatigue from being contacted by the school. There's automated messages that come from the School Board, there's calls from the school, and so you might not have as much feedback on a second survey as you would on the first one.''}
%
%
Engagement fatigue was present in the previous community engagement process. P4 said several rounds of proposals and negative feedback extended the process, which ultimately took more than a year.  \emph{``We kept taking surveys essentially over and over again''} (P2).
%

Though the engagement process was long, participants said the community still did not receive enough time during each iteration to digest possible boundary changes before being asked to share their feedback. 
P5 said, \emph{``They weren't releasing the information until the meeting where there were 300-plus people in a large cafeteria. They're handing out a printout of a PowerPoint and expecting people to be able to digest that and then provide feedback in the same meeting. In order to provide thoughtful, meaningful feedback, you need more time.''}
%
%
In response to the problem of not having enough time to digest information, P4 suggested having \emph{``overall access''} to the BoundarEase platform, so people would have enough \emph{``time to play with it, to look at it, to interact with it, to talk to a neighbor [about it].''}
P5 recommended leveraging the parent working group to create smaller community sessions, which could serve as \emph{``more neutral environments to explore the tool and the information, and to genuinely ask questions without fear of your neighbors feeling the opposite.''} 

\subsubsection{Interface Usability and Accessibility}

\paragraph{Current Strengths} 

Participants expressed an appreciation for the platform's brevity while still providing adequate, relevant information. P11 mentioned that \emph{``you don't want folks to lose interest or feel like they're overwhelmed by the quantity of content. So I think it's concise enough that it doesn't do that, but it does provide you enough information.''} 
%
BoundarEase generally appeared more understandable compared to the current process, despite it being a novel platform with new visualizations and interactions. P1 reflected holistically on the platform, stating, \emph{``The way the information is shared makes sense so that an average parent would be able to look at it and easily understand the impact of the change and what they were being asked to give feedback on.''} 
%
%
%
Participants specifically praised BoundarEase's multi-modal representation of the data---textual, numerical, and visual--- saying it made the data more accessible. P11 said, \emph{``I like the written explanations with the graphic. I think that's really nice that every written explanation has [a graphic] where needed.''} When talking about one of the visualizations, P5 called out their own preferences and acknowledged how other people might approach the data: \emph{``This is helpful for people who are more visual in that way. I tend to digest information through text better than graphics. But I do think that this is a really interesting way to show the feeder patterns.''}
%
%
Another strength was how responsive and interactive the platform is, changing based on rankings, addresses, and dropdown selections. P8 reflected, \emph{``I like how you can just kind of toggle back and forth between the different scenarios very easily with current boundaries versus new boundaries. It's so much easier than looking at multiple graphs across, you know, six different PowerPoint slides. It's just so much easier to get a snapshot of the data.''}  
Certain text descriptions and visualizations also received overwhelmingly positive reactions. Though it often took them a few moments to digest, participants found the feeder pattern visualization effective. P8 had \emph{``never seen anything like [it]''} but she found it \emph{``a great way to be able to see what's happening and where everyone's going.''} 
Participants also appreciated the detail and readability of the maps that showed the school boundary lines. The street-level view and the markers helped orient users: \emph{``The ability to just really quickly...figure out exactly what's going on here and get straight to your own street and actually see your house plotted on there with the address is really helpful''} (P6).
%

\paragraph{Improvement Suggestions} 

We also received various ideas to improve BoundarEase. In terms of visualizations, six participants found the radar chart for utilization confusing or offering little value. There were also suggestions on wording and additional statistics that would make the text descriptions more insightful. For example, P1 said the data on home-to-school distance obscured the potential effects on some families: \emph{``It's not really showing like an average increase travel time...If you're telling me only 8\% of students are gonna have an increase in this, but it's 30 minutes on average for this 8\% of students. That's a pretty big impact, whereas five minutes maybe isn't as big of an impact.''}
Participants suggested making it clearer that this is an exploratory tool, after learning that they could go back and change their inputs on the first page. P6 mentioned that \emph{``it'd be interesting to go back and...change the rankings and see how it all shakes out.''} 

\section{Discussion}

Unlike many other policymaking contexts, a school district rezoning often has a clear, direct, and immediate impact on families living within a district's boundaries. For families, this not only requires several individual-level considerations (perceived school quality, school usage, travel time), but it also requires them to consider those effects on other families. Engaging in policymaking at the school district level may seem like an easy entry into civic participation, but encouraging system-level discourse between parents and school districts has proved challenging and time-consuming. This often means many families do not have the time to partake in the process and, in turn, the final policies reflect the preferences of more privileged families who can partake. 

Given this backdrop, our research explores whether a data- and/or community-driven platform can help parents quickly and clearly formulate individual preferences based on data, while getting better visibility of what kind of system-level impact their preferences would have---and whether that plays a part in shaping their preferences. 

\subsection{Exploring BoundarEase's Ability to Help Users Connect Individual Policy Preferences to System-Wide Effects}

Recall our design questions:

\begin{enumerate}
    \item (DQ-1) How does an interface that depicts various school rezoning scenarios, with a focus on its impact on the district as a whole, prompt families in that district to consider the effect of their policy preferences on other families?
    \item (DQ-2) When exposed to individual stories of families who are adversely affected by a policy priority, how do families who support those policies react to their role in creating a system that hurts another family? 
    \item (DQ-3) If the district presents its rezoning goals with a shared vocabulary or set of community values, how do families perceive the district's intentions?
\end{enumerate}

For DQ-1, BoundarEase appeared to prompt participants to not only consider how proposed boundaries might impact their own families but also others.  Without prompting from researchers, several participants gamed out how a boundary policy might play out for families with students living in different parts of the district.  However, we found the platform had a limited impact on helping participants simultaneously consider how their individual preferences produced unfavorable outcomes for others.  With regards to DQ-2, we observed mixed feedback on the perspective-getting stories.  Some participants said exposure to other perspectives might add healthy frictions into the feedback process, but others wanted a more balanced sampling of perspectives---and some skipped over these stories entirely.  Arguably, we observed the most promising results for DQ-3. Participants appreciated how clearly the platform presented expected rezoning impacts, framed around the four pillars.  Several participants indicated that the tool's data-driven focus on communicating the impacts of a boundary change could prompt more rational decision-making and reduce emotional tension.  Indeed, some commented that the district's investment in such tools inspired greater trust and hope because it demonstrated a commitment to bringing transparency to a status-quo process that currently lacks it.

Together, our findings validate the potential for a tool like BoundarEase to help facilitate socio-politically complex processes like school attendance boundary planning in ways that might advance more equitable student assignment policymaking.  Below, we explore a few thematic areas revealed by the user study and reflect on several ways in which our key audiences---education researchers, HCI researchers, and school district practitioners---might integrate and build on our findings.

\paragraph{Considering Others and the Broader Collective.}  There were several instances during the user study where participants demonstrated deep reflectiveness on their roles and approaches during the recent rezoning, with several acknowledging how they allowed themselves to get trapped in their own bubbles as the process intensified. While BoundarEase didn't transform individual psychology or collective culture---for example, leading families to indicate that they might, in the future, concede individual gains for the collective---it did appear to scaffold participants' exposure to realities (data + stories) outside of their own bubbles. This approach of exposing users to different perspectives differs from many in the HCI space~\cite{bowenMetroFutures20202023, kimCrowdsourcingPerspectivesPublic2019, nelimarkkaComparingThreeOnline2014, kripleanSupportingReflectivePublic2012} in that it not only shares others' beliefs---possibly leading to reconsideration or even moderation of opinions~\cite{kripleanSupportingReflectivePublic2012}---but also the impact on the collective community and (potential) lived experiences within it. The latter may prove to be particularly beneficial to the discourse around school boundaries. Sharing personal stories has previously proven to be effective in promoting respect for the sharer's rationale, especially when the narratives involved suffering or possible suffering~\cite{kubinPersonalExperiencesBridge2021}. Further still, the exchange of narratives has been shown to reduce prejudice against outgroups and increase receptivity to persuasion~\cite{kallaReducingExclusionaryAttitudes2020}. These practices may help lay the groundwork for more collectivist-leaning perspectives.


\paragraph{Balancing Exploration with the Need for Structure.} A key takeaway from the user study was that participants appreciated the ability to explore and interact with the data to learn about the impact of a set of boundaries. Previous work aligns with feedback we received in that a ``learn-by-doing model'', combined with personalization and immediate responsiveness, was effective in strengthening participant knowledge--- one of the six key attributes in democratic deliberation~\cite{haaslyonsExploringRegionalFutures2014}. Many participants expressed excitement over being able to simulate different scenarios by reordering the pillars and seeing how that altered the projected impact. Some wanted to input addresses besides their own to get personalized information on how a set of boundaries would affect others. 

Importantly, this exploration and simulation is not unchecked, but rather structured through the finite permutations of the four pillars, the ability to add statistics to open-ended feedback, and the selectivity of the perspective-getting feature (as opposed to full access to all participant feedback). The latter two in particular strike a balance between expressiveness when giving feedback, and ``redundancy''~\cite{kripleanSupportingReflectivePublic2012}, which provides participants an opportunity to incorporate talking points and ideas from other community members. Not only may this reduce inflammatory language \cite{kripleanSupportingReflectivePublic2012}, but it can also support district analysis later in the policy planning process: administrators might more easily code feedback, or better understand what opinions the community converges on. We must be wary, though, of such a feature potentially leading to the oversimplification or disregard of less popular opinions, causing ``tyranny of the majority''~\cite{baumerCourseItPolitical2022}. Ultimately, BoundarEase tries to balance open-ended exploration with intentionally placed constraints.

\paragraph{Trust Through Improved Communication and Transparency.}  
Participants consistently reported that BoundarEase's clear presentation of data helped them understand and contextualize the district’s proposals and their potential impacts.
This increased accessibility is crucial for effective communication in community engagement processes~\cite{corbettProblemCommunityEngagement2018}. 
Greater accessibility enhances transparency~\cite{dickinson2019cavalry}, allowing community members to easily track each boundary iteration.  
Providing data on the four policy pillars also defines measurable outcomes, which can sustain public interest~\cite{saldivarCivicTechnologySocial2019} by enabling ongoing evaluation of boundary proposals.   
However, transparency alone does not ensure accountability without a well-structured engagement process~\cite{foxUncertainRelationshipTransparency2007}.
The design principles for crowdsourced policymaking proposed by~\cite{aitamurtoFiveDesignPrinciples2015}, including process modularity, clear goals and structure, continuous contact with the public, and closing the feedback loop at each stage of the process, are essential.
The BoundarEase platform supports modularity since it can be deployed across multiple boundary iterations, promoting consistency while permitting multiple entry points for feedback. 

While trust is not a prerequisite for accountability, it is crucial for fostering a shared understanding between decision-makers and the public~\cite{dickinson2019cavalry}.
Our studies revealed a general lack of trust in the school district. While this shared distrust galvanized some parent groups, it could pose a challenge around the \textit{trustworthiness} of information-sharing platforms like BoundarEase that districts might adopt in their engagement efforts~\cite{kripleanSupportingReflectivePublic2012}.
Furthermore, digital interfaces that replace relationships with efficient and transactional interactions can increase distrust by creating distance~\cite{corbettProblemCommunityEngagement2018, corbettRemovingBarriersCreating2019}.
To build trust, bi-directional information flows are essential~\cite{hardingHCICivicEngagement2015}.
BoundarEase facilitates this flow for one step of the boundary planning process by allowing decision-makers to share boundary decisions with justifications backed by data and enabling the community to provide feedback using the same data. 
Future work would involve extending the bi-directional flow throughout the entire engagement process.
Additionally, fostering relationships can build trust~\cite{corbettRemovingBarriersCreating2019, dickinson2019cavalry}.
For example, future iterations of BoundarEase can empower residents with information to mobilize others, tapping into community assets and wisdom \cite{dickinson2019cavalry}---though such information could also come with new risks, which we reflect on below.

\subsection{Situating BoundarEase for CSCW and Civic Technology Researchers}

We believe an important contribution of this project to CSCW and civic technology researchers is the introduction of a sociotechnical system to help facilitate a process that occurs frequently, affects virtually all US families whose children attend public schools, and is highly polarizing and contentious---but rarely explored in HCI or other academic disciplines.  Student assignment planning offers an academically complex and socially consequential domain for CSCW researchers to explore in future work.  The design of the BoundarEase system also reflects existing theories and frameworks of designing systems for public participation, which may further enable CSCW researchers coming from other domains to determine how to engage and advance future work.  For example, on Arnstein's ladder of citizen participation, BoundarEase may help advance existing district engagement practices from more token practices, like informing and consultation, towards distributing more power to community residents in ways that enable genuine participation by using the expected, concrete impacts of a policy change to inform refined proposals~\cite{arnsteinLadderCitizenParticipation1969}.  Similarly, on the spectrum of public participation, it may help move community input towards genuine involvement and collaboration \cite{nelimarkkaComparingThreeOnline2014}.  Achieving this movement will undoubtedly require updates to BoundarEase that include multiple feedback loops, and a transparent process by which districts incorporate feedback to inform future proposals.  Doing this fully and in a way that helps advance equitable student assignment policies, however, will also require ensuring that subgroups have an equal voice in such participation---which is not guaranteed given typical imbalances in public participation \cite{goochAmplifyingQuietVoices2018}.  We return to these topics in the limitations and future work section below.

\subsection{Expanding Field Deployments}
\subsubsection{Integrating BoundarEase into School Districts}
With minimal modifications, BoundarEase may offer practical improvements in status quo community engagement practices around boundary planning across many US public school districts.  We have already demoed the platform to one other district not involved in this study and received tremendous interest.  One reason for this interest may be that many districts already have pillars that govern their rezoning processes, and thus, may see BoundarEase as a structured system for foregrounding those pillars and the impacts proposed policies might have on families.  As discussed throughout this paper, such structure may help advance transparency and trust in a community process that is typically mired in controversy, contention, and oftentimes, irrationality---especially if it's clear how one's preferences over such pillars are factored into the boundary redrawing process.  This may be particularly useful early on in this process when the objective is to obtain a broader understanding of community sentiment toward a wide range of proposals.  The presence of such a system may also enable parent groups and other civil society actors beyond district leadership to highlight how existing boundaries adversely impact certain families over others, and generate hypothetical new boundaries to support change advocacy efforts.

In addition to primarily asynchronous use cases, districts may adapt BoundarEase to settings of ``augmented deliberation'' \cite{gordonAugmentedDeliberationMerging2011}---e.g., as a tool for helping to facilitate synchronous community feedback sessions, which often devolve into shouting matches.  This could be particularly valuable, as the synchronous setting may create natural opportunities for elaboration \cite{aitamurtoDisagreementAgreementElaboration2023} when community members disagree on the proposals, impacts, and justifications surfaced through BoundarEase---which may, in turn, yield a more productive engagement process.

\subsubsection{Connecting BoundarEase to Other Domains}

From our user study, we demonstrated BoundarEase's potential to help community members understand and provide feedback on boundary changes. 
The platform can be extended to diverse domains, such as urban planning and policymaking, enriching community engagement and decision-making processes in sociotechnical systems.
In these contexts, decision-making is characterized by prolonged timelines, numerous stakeholders, and complex bureaucracy~\cite{rittel1984planning}.
BoundarEase's unique features, including text-based summaries and interactive visualizations, contribute to an efficient and holistic understanding of individual and collective impacts of potential policy changes.
The scaffolded feedback mechanisms provide structured and streamlined ways of gathering community input, which can be applied to other civic use cases.  
Moreover, the platform's potential extends to instances where there is ineffective communication between constituents and decision-makers, such as constituent correspondence in Congress~\cite{mcdonaldConstituentCorrespondenceCongress}.
The reconfigurability of BoundarEase, allowing community members to create different versions of algorithmically generated boundaries, aligns with the emerging field of democratic artificial intelligence (AI), in which AI tools optimize for human preferences, generating policies that better reflect the diverse needs of communities~\cite{kosterHumancentredMechanismDesign2022}.

\subsection{Risks and Ethical Considerations}
While several participants noted BoundarEase could reduce barriers for some families to have their voices heard, we note that a tool that reduces information frictions may also serve as a ``power tool'' for already engaged families.  If they are no longer pouring into creating bespoke summaries that seek to aggregate and present the information BoundarEase presented---as many indicated they were during the rezoning process we studied---they may have more time for activating parents and lobbying the school board to further their agendas.  This could prompt practices like ``super-posting'', where discussion spaces are flooded with perspectives representing a powerful minority \cite{macneilFindingPlaceDesign2021}, intensifying support for agendas that may counter those seeking to advance more equitable access to quality education.  While recent advances in natural language processing may make it easier for district officials to efficiently detect practices like super-posting in large feedback corpora \cite{beefermanFeedbackMap2023}, we anticipate that cultivating community engagement with norms that effectively discourage such practices will continue to be an open challenge for districts, and perhaps, a critical opportunity for innovation with the support of HCI researchers and civic designers.

\subsection{Limitations and Future Work}
There are several limitations of our study, which serve as important bases for future work.  One involves its sandboxed nature.  In the future, testing BoundarEase through a field deployment can help offer a more complete view of how such a system might help foster more productive boundary planning processes, with more equitable outcomes---and what might block such progress.  Conducting field deployments across multiple school districts may also shed light on the generalizability of our preliminary findings.  Important for a field deployment will also be the design and iteration of multiple feedback loops that incorporate community feedback to inform new boundary proposals---another limitation of the present study.  Such feedback loops may incorporate additional features that help people understand where they stand in relation to others, which could help advance transparency, create healthy points of reflection and ``friction'' \cite{kornCreatingFrictionInfrastructuring2015}, and perhaps even help drive conversations towards consensus \cite{liuConsensUsSupportingMultiCriteria2018}.  

All of these sociotechnical considerations, of course, hinge on a critical question: \textit{whose} voices are represented in community engagement processes?  While our study included the voices of community residents with different rezoning priorities (which may reflect their broader ideological viewpoints), most appeared to be actively engaged parents from relatively affluent backgrounds, with several participating in the parent working group created by the district.  These were largely parent leaders---those who other, less involved parents looked to for information and guidance.  On the one hand, this dynamic of ``leaders'' and ``followers'' is the norm in civic engagement, as evidenced by the two-step theory of communication~\cite{lazarsfeld1968people}, and even more recently proposed paradigms like liquid democracy \cite{blum2016liquid}.  On the other hand, one of the promises of sociotechnical systems for civic engagement is that they can ``meet people where they are'' to create new on-ramps to informed community engagement---regardless of the time pressures, socioeconomic constraints, or other obstacles they might face.  While it may sound aspirational, this is an important design direction for future iterations of BoundarEase---particularly those that are deployed in the field: how to enact design principles like supporting multiple languages \cite{aitamurtoFiveDesignPrinciples2015}, technological accessibility for those with lower levels of digital literacy \cite{yavuzDigitalDivideCitizen2023}, and reaching underrepresented groups through community members they trust to inspire genuine participation \cite{le-dantec2015strangers}. Focusing on technological advancements to increase participation may disrupt existing human-led methods of ensuring equity in civic decision-making processes~\cite{ledantec2015planning}. Indeed, both technical and human infrastructure will be required to ensure that historically underrepresented voices are heard and meaningfully influence the policymaking process.

\section{Conclusion}

In this paper, we introduce BoundarEase, a sociotechnical system for fostering more constructive community engagement around changing school attendance boundaries.  Attendance boundaries act as a consequential policy lever that thousands of school districts across the US use to shape which schools students have access to.  Through a formative user study with 16 community members in a large public US school district, we identify several frictions in existing community engagement practices, including ones that threaten to hamper the development of more equitable student assignment policies.  We design BoundarEase to try and mitigate these frictions---namely, by clarifying the impacts that policy changes might have on a given family, but also, families who are different from them---to foster more collective consideration during a process that is otherwise characterized by a lack of transparency and rampant individualism.  We find that BoundarEase reduces barriers to understanding the intricacies of how new attendance boundaries might impact families, and as a part of this, also prompts families to reflect on how others (beyond themselves) might be affected.  As it stands today, the BoundarEase system may serve as a useful resource for school districts seeking to facilitate engagement around attendance boundary changes; however, its present design offers several opportunities for future work.  These opportunities include exploring how the platform might evolve to support policy iteration by factoring in community feedback; how it might be used in synchronous dialogue settings to help facilitate and moderate otherwise contentious community feedback processes; and, crucially, which additional technological and human infrastructure is required to ensure that it does not simply serve as a power tool for the already-powerful, and instead, truly levels existing informational inequities in local education policymaking.  

School attendance boundary planning is a highly controversial, yet critically important lever for advancing equitable life outcomes for students and families---particularly those from historically disadvantaged groups.  We hope our study offers a proof of concept of a sociotechnical system to help facilitate constructive community engagement on this socially-consequential topic, as well as a new challenge domain that is ripe for CSCW researchers to critically, and meaningfully, advance.

\begin{acks}
We thank everyone who tested out and provided feedback on BoundarEase.
Special thanks to Claire Schuch and Rebecca Eynon for insightful discussions and support.
We thank the reviewers for their helpful comments and suggestions.
The first author was supported by the National Science Foundation Graduate Research Fellowship Program under Grant No. 2141064.
This research was also supported by Northeastern University. Any opinions, findings, and conclusions or recommendations expressed in this material are those of the authors and do not necessarily reflect the views of the National Science Foundation.
\end{acks}

\bibliographystyle{ACM-Reference-Format}
\bibliography{references}

\newpage
\appendix

\section{Appendix}


\subsection{Four Pillars}

Table~\ref{tab:pillardefinitions} contains definitions of the four policy pillars: socioeconomic (SES) diversity, home-to-school distance, feeder patterns, and utilization.

\begin{table}[t]
    \centering
    \caption{Definitions of the four policy pillars.}
    \begin{tabular}{@{}p{0.17\linewidth}|p{0.78\linewidth}}
        \toprule
        \textbf{Pillar}&\textbf{Definition}\\
        \midrule
        Feeder patterns& The neighborhood schools students are zoned for. A split feeder pattern occurs when students from one school are zoned to multiple schools after completing the highest grade level (e.g. one middle school feeds into two high schools). \\
        Home-to-school distance& The distance from home to a school in minutes\\
        SES diversity& Diversity in students' socioeconomic status (i.e. low, medium, or high)\\
        Utilization& The percentage of a school's capacity occupied by students and staff\\
        \bottomrule
    \end{tabular}
    \label{tab:pillardefinitions}
\end{table}

\subsection{Formative Study Participants}

Table~\ref{tab:formativeparticipants} contains information about each participant in the formative study.

\begin{table}[]
    \centering
    \caption{Information about each participant in the formative study. 
    ``SES'' refers to the dominant socioeconomic status (low, medium / med, or high) of students in the participants' neighborhoods.
    ``Children's Grades'' refers to a list of academic grades (from preschool to 12th grade) a participant's children are in. 
    ``School Count'' describes the number of schools their children attended at the time of the interview. 
    ``Involvement'' summarizes how each participant described their engagement with the boundary planning process.}
    \begin{tabular}{@{}p{0.11\linewidth}|p{0.07\linewidth}|p{0.05\linewidth}|p{0.10\linewidth}|p{0.065\linewidth}|p{0.5\linewidth}}
        \toprule
        Participant&Gender&SES&Children's Grades&School Count&Involvement\\
        \midrule
        F1&Female&High&9, 12&1&Active: attended most community and school board meetings\\
        F2&Female&High&6, 8, 10&2&Active: attended community meetings and \newline involved in Parent Teacher Student Organization\\
        F3&Female&High&4, 8&2&Active: part of the parent working group\\
        F4&Male&High&7, 9, 11&2&Active: part of the parent working group\\
        F5&Female&High&1, 8, 9&3&Less active: attended initial community meetings but stopped; filled out some surveys\\
        F6&Female&Med&7, 10&2&Active: attended most community meetings and filled out surveys\\
        F7&Female&High&7, 10&2&Active: attended most community and school board meetings\\
        F8&Female&High&4, 8&2&Moderately active: attended some community meetings and filled out a survey\\
        F9&Female&Med&1, 4, 6&2&Active: part of the parent working group\\
        F10&Female&High&10&1&Active: part of the parent working group\\
        F11&Female&High&5, 9&2&Moderately active: attended four community meetings\\
        F12&Female&High&6, 9&2&Moderately active: attended some online \newline community meetings and one school meeting\\
        F13&Female&Med&pre-K, 6, 7&2&Less active: didn't attend many events related to the rezoning; involved in curriculum discussions\\
        F14&Female&Med&2, 7&1&Less active: listened to one online community meeting and filled out some surveys\\
        F15&Female&Med&5, 8&2&Active: part of the parent working group\\
        F16&Female&High&1, 4, 6&2&Less active: filled out some surveys and listened to a few meeting recordings\\
        \bottomrule
    \end{tabular}
    \label{tab:formativeparticipants}
\end{table}

\subsection{BoundarEase System}\label{appendix:system}

\subsubsection{Start Page}

\begin{figure}[h]
\includegraphics[width=0.8\linewidth]{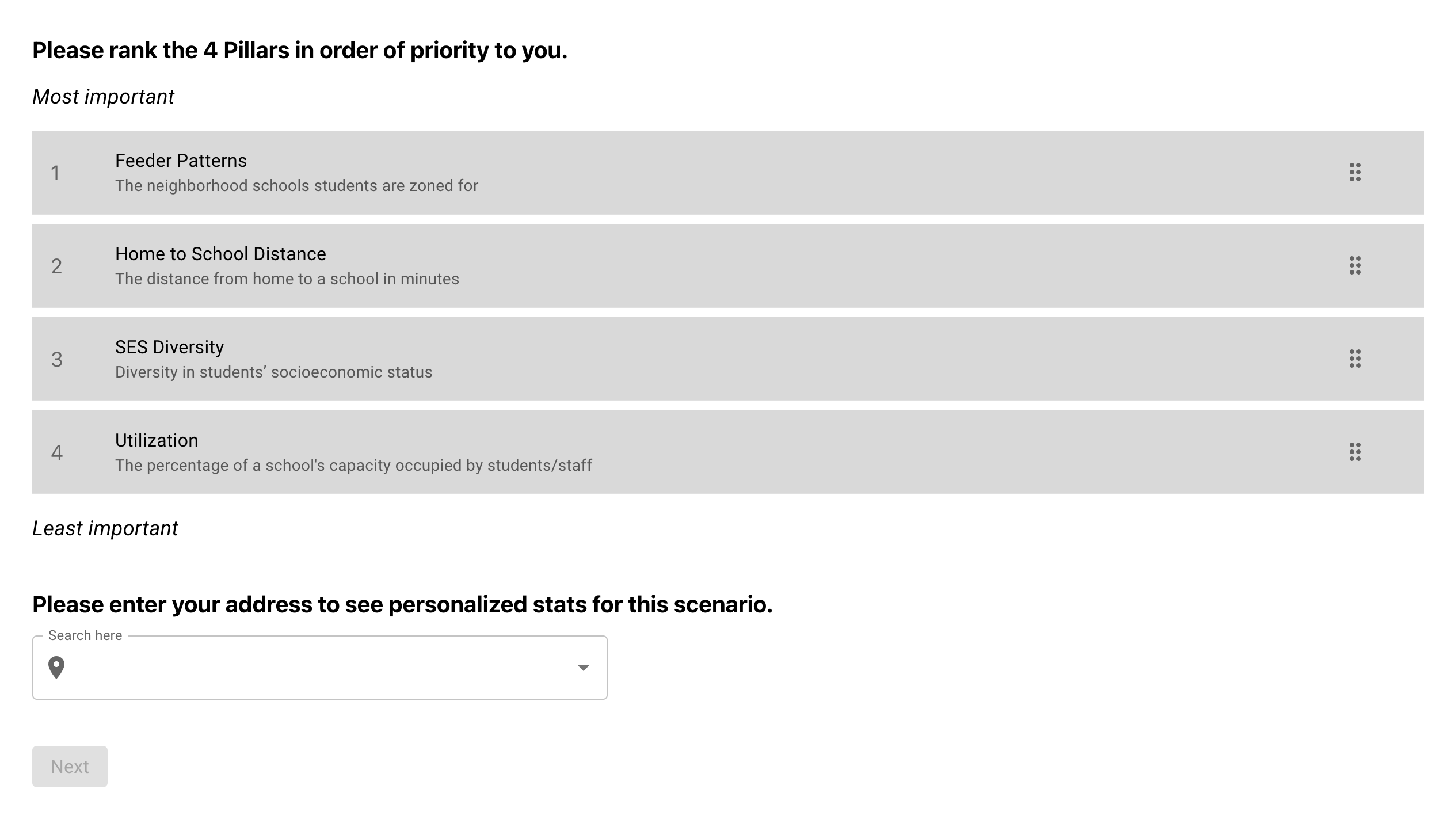}
\caption{BoundarEase's start page. The pillar cards can be dragged in the order of importance for the user. }
\label{fig:startpage}
\Description{The BoundarEase start page. At the top is the text ''Please rank the 4 Pillars in order of priority to you.'' Below that is the text ''Most important,'' then four vertically-stacked draggable cards, and then the text ''Least important.'' Each card has a number, one through four, on the leftmost side that changes based on where it's dragged. It also has the name and a short definition of the pillar. On the rightmost side of the card is a drag icon, two columns of three dots each. Beneath this ranking section is the text ''Please enter your address to see personalized stats for this scenario'' and an autocomplete input for searching and entering an address. Finally, there is a disabled Next button.}
\end{figure}

Fig.~\ref{fig:startpage} shows a screenshot of the first page.

\subsubsection{Text for SES Diversity Pillar}

\begin{figure}[h]
\centering
\includegraphics[width=0.8\linewidth]{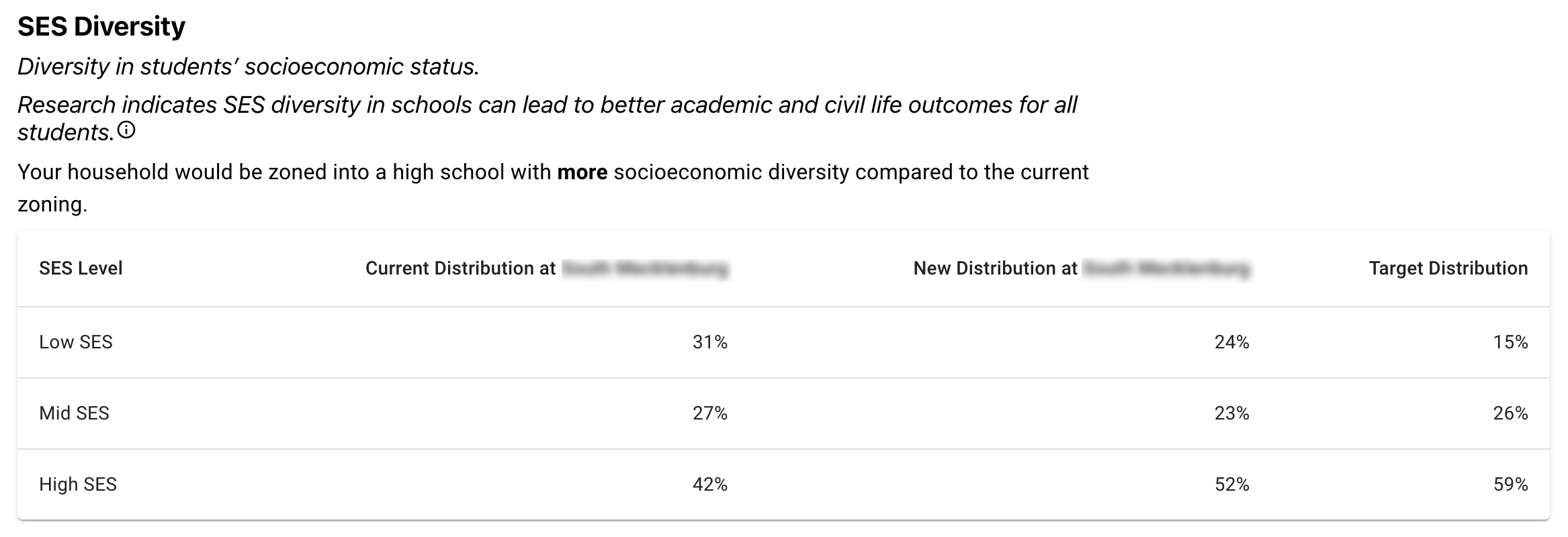} 
\caption{Text explaining the impact on the individual user, and a table showing three SES distributions--- the school they are currently zoned for, the school they would be zoned for given the proposed boundaries, and the district's target distribution for any school--- across ``low,'' ``mid,'' and ``high'' categories. The target distribution remains the same across schools and proposals.}
\label{fig:sesdiversitypersonal}
\Description{The SES Diversity text and table. There is the name of the pillar followed by its definition. Below that is the additional text ''Research indicates SES diversity in schools can lead to better academic and civil life outcomes for all students'' with an information icon. Then there is a personalized sentence ''Your household would be zoned into a high school with more socioeconomic diversity compared to the current zoning.'' Beneath that is a table with four columns. The leftmost column has the header ''SES Level'' and the rows ''Low SES'', ''Mid SES'', and ''High SES''. The remaining column headers are ''Current Distribution at [the name of the user's current school]'', ''New Distribution at [the name of the user's school with the proposed boundaries], and ''Target Distribution''. The rows for each column are the percentages corresponding to each SES category.}
\end{figure}

First, data specific to the user is displayed; the dynamic text states: \emph{``Your household would be zoned into a high school with \_\_\_ socioeconomic diversity compared to the current zoning,''} with the blank being filled as \emph{``more''} or \emph{``less''} accordingly. A table (Fig.~\ref{fig:sesdiversitypersonal}) shows the percentages of students' SES in schools relevant to the user, as well as the district's SES distribution goal. 

\subsubsection{Text for Feeder Pattern Pillar}

For this pillar, the text describes the user's current feeder pattern and their new feeder pattern with the proposed boundaries. Following this, we describe the district-wide impact as: \emph{``Overall, XXX (X\%) students would experience split feeder patterns, which is \_\_\_ than the YYY (Y\%) students who experience split feeder patterns with the current boundaries,''} with the blank being \emph{``higher''} or \emph{``lower.''} The sentence after follows the same format but for intact feeder patterns.

\subsubsection{Text for Utilization Pillar}

We include the district's definitions of utilization in addition to the name and general definition. The dynamic text highlights whether the user's current school is within target enrollment (i.e. 100\%, though 130\% is possible as an absolute maximum) and the actual number of enrolled students, saying: \emph{``Your currently zoned school ([school name]) is not within target enrollment, since XXX students are enrolled (X\% utilization) and YYY students is 100\% utilization.''} Following this is a nearly identical sentence describing the utilization of the new school given the proposed boundaries. To give insight into the district-wide impact, we share the number of schools that are currently within target enrollment and how many would be within target enrollment given the proposed boundaries.

\subsection{User Study Participants}

Table~\ref{tab:userparticipants} contains information about each participant in the user study.

\begin{table}[]
    \centering
    \caption{Information about each participant in the user study. 
    ``SES'' refers to the dominant socioeconomic status (low, medium / med, or high) of students in the participants' neighborhoods.
    ``Children's Grades'' refers to a list of academic grades (from preschool to 12th grade) a participant's children are in. 
    ``School Count'' describes the number of schools their children attended at the time of the interview. 
    ``Involvement'' summarizes how each participant described their engagement with the boundary planning process.}
    \begin{tabular}{@{}p{0.11\linewidth}|p{0.07\linewidth}|p{0.05\linewidth}|p{0.10\linewidth}|p{0.065\linewidth}|p{0.5\linewidth}}
        \toprule
        Participant&Gender&SES&Children's Grades&School Count&Involvement\\
        \midrule
        P1&Female&High&3&1&Less active: found it hard to stay informed about the process; PTA president provided information\\
        P2&Female&High&6, 10&2&Active: part of the parent working group\\
        P3&Female&Med&3, 6&2&Active: part of the parent working group\\
        P4 (F1)&Female&High&9, 12&1&Active: attended most community and school board meetings\\
        P5&Female&High&pre-K, 1&1&Active: part of the parent working group\\
        P6 (F4)&Male&High&7, 9, 11&2&Active: part of the parent working group\\
        P7 (F2)&Female&High&6, 8, 10&2&Active: attended community meetings and \newline involved in Parent Teacher Student Organization\\
        P8 (F9)&Female&Med&1, 4, 6&2&Active: part of the parent working group\\
        P9&Female&Med&1&1&Moderately active: attended some community meetings\\
        P10&Female&High&3, 6&2&Active: attended most community meetings\\
        P11 (F6)&Female&Med&7, 10&2&Active: attended most community meetings and filled out surveys\\
        P12 (F5)&Female&High&1, 8, 9&3&Less active: attended initial community meetings but stopped; filled out some surveys\\
        \bottomrule
    \end{tabular}
    \label{tab:userparticipants}
\end{table}

\end{document}